\documentclass[a4paper,10pt]{revtex4}
\usepackage{mathrsfs}
\usepackage{graphicx}
\usepackage{latexsym}
\usepackage{amsmath}
\usepackage{amssymb}
\usepackage{textcomp}
\usepackage{amsbsy}
\usepackage{graphics}
\usepackage{epstopdf}
\usepackage{color}

\begin{document}

\tolerance=5000

\title{Towards a smooth unification from an ekpyrotic bounce to the dark energy era}

\author{Shin'ichi~Nojiri$^{1,2}$\,\thanks{nojiri@gravity.phys.nagoya-u.ac.jp},
Sergei~D.~Odintsov$^{3,4}$\,\thanks{odintsov@ieec.uab.es},
Tanmoy~Paul$^{5,6}$\,\thanks{pul.tnmy9@gmail.com}} \affiliation{
$^{1)}$ Department of Physics, Nagoya University,
Nagoya 464-8602, Japan \\
$^{2)}$ Kobayashi-Maskawa Institute for the Origin of Particles
and the Universe, Nagoya University, Nagoya 464-8602, Japan \\
$^{3)}$ ICREA, Passeig Luis Companys, 23, 08010 Barcelona, Spain\\
$^{4)}$ Institute of Space Sciences (IEEC-CSIC) C. Can Magrans
s/n, 08193 Barcelona, Spain\\
$^{5)}$ Department of Physics, Chandernagore College, Hooghly - 712 136, India\\
$^{6)}$ Labaratory for Theoretical Cosmology, International Centre of Gravity and Cosmos,
Tomsk State University of Control Systems and Radioelectronics (TUSUR), 634050 Tomsk, Russia}


\tolerance=5000

\begin{abstract}
In the context of a ghost free $f(R,\mathcal{G})$ model, we present a non-singular cosmological scenario in which the universe initially contracts 
through an ekpyrotic phase having a bouncing like behaviour, and following the bounce, it smoothly transits to a matter or radiation like deceleration 
era which is further smoothly connected to the dark energy era at present epoch. The ghost free character of the model is ensured by the presence of a 
Lagrange multiplier, and we consider the Gauss-Bonnet (GB) coupling function in such a way that it gets compatible with the event GW170817. Using 
suitable reconstruction technique, we obtain the non-trivial scalar field potential as well as the GB coupling function. Such scalar potential and 
GB coupling function source a smooth unified scenario from an ekpyrotic bounce to the dark energy era with an intermediate deceleration era. The 
occurrence of ekpyrotic contraction phase justifies the resolution of the anisotropic problem (also known as BKL instability) in the background evolution. 
Consequently we determine the background Hubble parameter and the corresponding 
effective equation of state parameter, and discussed several qualitative features 
of the model. The Hubble radius shows an asymmetric behaviour around the bounce, in particular, the evolution of the Hubble radius leads to the generation era 
of the primordial perturbation modes far before the bounce in the deep sub-Hubble regime. Accordingly we perform the scalar and tensor perturbation 
evolution in the present context, and as a result, the scalar power spectrum at large scale modes is found to be problematic. 
Thus an extended scenario is proposed where we consider a pre-ekpyrotic phase having the equation of state parameter is less than unity, 
and re-examine the scalar and tensor power spectra, on large scales that cross the Hubble radius during the pre-ekpyrotic stage. 
In this regard, the GB coupling function shows considerable effects in reducing the tensor to scalar ratio compared to the case where the GB coupling 
is absent. Furthermore the dark energy epoch is consistent with the Planck+SNe+BAO data.
\end{abstract}


\maketitle

\section{Introduction}
One of the key challenge in modern cosmology is to ascertain whether the universe started its expansion from Big Bang singularity or from a bouncing 
like stage that is free of singularity. Inflation is an appealing early universe scenario as it resolves the flatness and horizon problems, and most 
importantly, produces a scale invariant primordial power spectrum which seems  to be consistent with the Planck data 
\cite{Guth:1980zm,Linde:2005ht,Langlois:2004de,Riotto:2002yw,Baumann:2009ds}. However the inflation model(s), 
if we go backward in time, are plagued with a curvature singularity popularly known as Big Bang singularity. Probably, a quantum theory of gravity 
will be able to avoid this singularity. However in absence of a fully accepted quantum gravity, bouncing cosmology is the most promising scenario 
that can lead to a non-singular universe 
\cite{Brandenberger:2012zb,Brandenberger:2016vhg,Battefeld:2014uga,Novello:2008ra,Cai:2014bea,deHaro:2015wda,Lehners:2011kr,Lehners:2008vx,
Cai:2016hea,Li:2014era,Brizuela:2009nk,Cai:2013kja,Quintin:2014oea,Cai:2013vm,Raveendran:2017vfx,Raveendran:2018yyh,Raveendran:2018why,
Koehn:2015vvy,Odintsov:2015zza,Koehn:2013upa,Battarra:2014kga,Martin:2001ue,Khoury:2001wf,
Buchbinder:2007ad,Brown:2004cs,Hackworth:2004xb,Peter:2002cn,Gasperini:2003pb,Creminelli:2004jg,Lehners:2015mra,
Mielczarek:2010ga,Lehners:2013cka,Cai:2014xxa,Cai:2007qw,Cai:2012va,Cai:2014zga,Avelino:2012ue,Barrow:2004ad,Haro:2015zda,Elizalde:2014uba,Banerjee:2020uil,
Das:2017jrl}.

Among various bounce models proposed so far, matter bounce model earned a lot of attention as it concomitantly describes a scale invariant power spectrum 
consistent with the observational data and a late matter dominated era during the expanding universe 
\cite{deHaro:2015wda,Cai:2008qw,Finelli:2001sr,Quintin:2014oea,Raveendran:2017vfx,Cai:2011ci,Cai:2013kja,
Brandenberger:2009yt,deHaro:2014kxa,
Qiu:2010ch,deHaro:2012xj,Elizalde:2019tee,Elizalde:2020zcb,Nojiri:2019lqw,WilsonEwing:2012pu}. However the matter bounce scenario (MBS) are hinged with some serious issues, like:
\begin{itemize}
\item The spacetime anisotropic energy density seems to grow faster than that of the bouncing agent during the contracting phase, which in turn makes the background 
evolution unstable (also known as BKL instability) \cite{new1}.
 
\item The Hubble radius in MBS monotonically increases with time after the bounce, i.e., the MBS is unable to explain the late time acceleration or 
equivalently the dark energy era of universe, which, 
in fact, is not consistent with the recent supernovae observations indicating a current accelerating stage of universe 
\cite{Perlmutter:1996ds,Perlmutter:1998np,Riess:1998cb}.
 
\item Beside these two problems from the perspective of background evolution, MBS generally predicts a large value of tensor to scalar ratio (compared 
to the Planck constraint \cite{Akrami:2018odb}) in the 
perturbation evolution, i.e., the scalar and tensor perturbations get comparable amplitudes to each other \cite{Brandenberger:2016vhg}. 
\end{itemize}

Attempts have been made to resolve these issues in the field of modified theories of gravity. However such attempts hardly explain 
all of the above issues concomitantly. For example, the articles \cite{Elizalde:2020zcb,Nojiri:2019lqw,Raveendran:2017vfx,Raveendran:2018why} 
proposed an extended matter bounce scenario in higher curvature models with the Lagrange multiplier term or in two scalar field models, 
where the third problem seems to be resolved, however the first and second problems persist. The article \cite{Raveendran:2018yyh} proposed an 
ekpyrotic bounce scenario in two scalar field models (where 
the curvature perturbation power spectrum gets almost scale invariant due to its interaction 
with the iso-curvature perturbation), which avoids the BKL instability, however suffers in explaining the dark energy issue. 
Moreover the bounce scenario in \cite{Odintsov:2020zct,Odintsov:2021yva} is able 
to resolve the second and third problems, however the first one persists, i.e., the models suffer from the BKL instability.

In the present paper, we propose a cosmological scenario which -- (1) smoothly unifies a non-singular bounce to a viable dark energy era, 
(2) free from the BKL instability and (3) predicts a tensor to scalar ratio that is indeed consistent with the Planck data, i.e, all the 
aforementioned issues are simultaneously addressed. In particular, the universe initially contracts through an ekpyrotic phase of contraction, and after 
the bounce, it smoothly transits to a matter or radiation like deceleration era which is further smoothly connected to the dark energy epoch. The occurrence 
of ekpyrotic phase justifies the resolution of anisotropic problem and makes the present unified scenario more natural compared to the previously discussed 
in \cite{Odintsov:2020zct,Odintsov:2021yva}. For the gravity theory, we consider $f(R,\mathcal{G})$ 
gravity which turns out to be a ghost free theory due to the presence of the Lagrange multiplier in the gravitational action, as developed in 
\cite{Nojiri:2018ouv} 
($R$ and $\mathcal{G}$ being the Ricci scalar and the Gauss-Bonnet scalar, respectively). The cosmology 
of $f(R,\mathcal{G})$ gravity from various perspectives have been discussed in 
\cite{Nojiri:2005vv,Li:2007jm,Carter:2005fu,Nojiri:2019dwl,Odintsov:2020sqy,Odintsov:2020zkl,
Bamba:2020qdj,Cognola:2006eg}. Furthermore the holographic correspondence of $f(R,\mathcal{G})$ 
gravity has been established in \cite{Nojiri:2020wmh}. 
In the present context, the $f(R,\mathcal{G})$ model gets compatible with the event GW170817 \cite{GBM:2017lvd}, 
according to which the speed of gravitational wave is unity, 
owing to certain choice of the Gauss-Bonnet (GB) coupling function. 
In effect, the GB coupling function is found to have significant contributions in making the tensor to scalar 
ratio consistent with the observational data. With such $f(R,\mathcal{G})$ model, 
we study the scalar and tensor perturbation power spectra for two different scenario depending on the initial conditions, particularly -- 
(1) in the first scenario, the universe undergoes through an ekpyrotic phase of contraction at distant past and consequently the 
perturbation modes generate during the same, while, (2) in the second scenario, the ekpyrotic phase is preceded by a pre-ekpyrotic stage having the equation 
of state parameter being less than unity, and thus the perturbation modes generate during the pre-ekpyrotic phase. In the second scenario, the pre-ekpyrotic 
stage is smoothly connected to the ekpyrotic one from the continuity of scale factor and the Hubble parameter at the junction point of time. 
Actually, the existence of such a pre-ekpyrotic phase is important in order to produce a scale invariant power spectrum at large scales 
in the present context. Detailed qualitative features are discussed at appropriate places of the paper.

The following notations will be used throughout the paper: $t$ is the cosmic time, $\eta$ is the conformal time defined by $\eta = \int\frac{dt}{a(t)}$ 
(with $a(t)$ being the scale factor of the universe), an overdot denotes $\frac{d}{dt}$ and an overprime represents $\frac{d}{d\eta}$. Moreover the 
conversions $1\,\mathrm{GeV} = 1.52\times10^{24}\mathrm{sec}^{-1}$ and $1\,\mathrm{By}^{-1} = \left(\frac{3.17}{1.52}\right)\times10^{-41}\mathrm{GeV}$ may 
be useful.

\section{Essential features of a ghost-free $f(R,\mathcal{G})$ gravity compatible with the GW170817 event\label{SecII}}

In this section we shall recall the essential features of the
ghost free $f(R,\mathcal{G})$ gravity theory developed in Ref.~\cite{Nojiri:2018ouv}. We consider
$f(R,\mathcal{G}) = \frac{R}{2\kappa^2} + f(\mathcal{G})$ which, owing to the presence of $f(\mathcal{G})$, contains ghosts with respect to
 perturbations of the spacetime metric. However the ghost modes may be eliminated by introducing a Lagrange multiplier $\lambda$ in
the standard $f(\mathcal{G})$ gravity action
\cite{Nojiri:2018ouv}, leading to a ghost-free action, as follows, 
\begin{equation}
\label{FRGBg19} S=\int d^4x\sqrt{-g} \left(\frac{1}{2\kappa^2}R 
+ \lambda \left( \frac{1}{2} \partial_\mu \chi \partial^\mu \chi 
+ \frac{\mu^4}{2} \right) - \frac{1}{2} \partial_\mu \chi \partial^\mu \chi
+ h\left( \chi \right) \mathcal{G} - V\left( \chi \right) + \mathcal{L}_\mathrm{matter}\right)\, ,
\end{equation}
where $\mu$ is a constant having mass dimension $[+1]$. 
Varying the action with respect to the Lagrange multiplier $\lambda$, one obtains the 
following constraint equation,
\begin{equation}
\label{FRGBg20} 
0=\frac{1}{2} \partial_\mu \chi \partial^\mu \chi + \frac{\mu^4}{2} \, .
\end{equation}
The kinetic term is effectively a constant, so it can be safely encapsulated within the scalar potential, as
\begin{equation}
\label{FRGBg21} 
\tilde V \left(\chi\right) \equiv \frac{1}{2}
\partial_\mu \chi \partial^\mu \chi + V \left( \chi \right)
= - \frac{\mu^4}{2} + V \left( \chi \right) \, ,
\end{equation}
and consequently the action of Eq.~(\ref{FRGBg19}) becomes
\begin{equation}
\label{FRGBg22} 
S=\int d^4x\sqrt{-g} \left(\frac{1}{2\kappa^2}R 
+ \lambda \left( \frac{1}{2} \partial_\mu \chi \partial^\mu \chi 
+ \frac{\mu^4}{2} \right) + h\left( \chi \right) \mathcal{G}
 - \tilde V\left( \chi \right) + \mathcal{L}_\mathrm{matter}\right) \, .
\end{equation}
The scalar and gravitational equations of motion for the action (\ref{FRGBg22}) take the  form
\begin{align}
\label{FRGBg23} 
0 =& - \frac{1}{\sqrt{-g}} \partial_\mu \left(
\lambda g^{\mu\nu}\sqrt{-g} \partial_\nu \chi \right)
+ h'\left( \chi \right) \mathcal{G} - {\tilde V}'\left( \chi \right) \, , \\
\label{FRGBg24} 
0 =& \frac{1}{2\kappa^2}\left(- R_{\mu\nu} 
+ \frac{1}{2}g_{\mu\nu} R\right) + \frac{1}{2} T_{\mathrm{matter}\,\mu\nu}
 - \frac{1}{2} \lambda \partial_\mu \chi \partial_\nu \chi
 - \frac{1}{2}g_{\mu\nu} \tilde V \left( \chi \right)
+ D_{\mu\nu}^{\ \ \tau\eta} \nabla_\tau \nabla_\eta h \left( \chi \right)\, ,
\end{align}
where $D_{\mu\nu}^{\ \ \tau\eta}$ is of the following form,
\begin{align}
D_{\mu\nu}^{\ \ \tau\eta}=&\left( \delta_{\mu}^{\ \tau}\delta_{\nu}^{\ \eta} + \delta_{\nu}^{\ \tau}\delta_{\mu}^{\ \eta} 
 - 2g_{\mu\nu}g^{\tau\eta} \right) R + \left( -4g^{\rho\tau}\delta_{\mu}^{\ \eta}\delta_{\nu}^{\ \sigma}
 - 4g^{\rho\tau}\delta_{\nu}^{\ \eta}\delta_{\mu}^{\ \sigma} + 4g_{\mu\nu}g^{\rho\tau}g^{\sigma\nu} \right) R_{\rho\sigma}\nonumber\\
&+4R_{\mu\nu}g^{\tau\eta} - 2R_{\rho\mu\sigma\nu} \left(g^{\rho\tau}g^{\sigma\nu} + g^{\rho\eta}g^{\sigma\tau}\right)
\nonumber
\end{align}
with having in mind $g^{\mu\nu}D_{\mu\nu}^{\ \ \tau\eta} = 4\left[-\frac{1}{2}g^{\tau\eta}R + R^{\tau\eta} \right]$. 
Upon multiplication of Eq.~(\ref{FRGBg24}) with $g^{\mu\nu}$, we get
\begin{equation}
\label{FRGBg24A} 
0 = \frac{R}{2\kappa^2} + \frac{1}{2}
T_\mathrm{matter} + \frac{\mu^4}{2} \lambda - 2 \tilde V \left(
\chi \right) - 4 \left( - R^{\tau\eta} + \frac{1}{2} g^{\tau\eta}
R \right) \nabla_\tau \nabla_\eta h \left( \chi \right) \, ,
\end{equation}
and solving  Eq.~(\ref{FRGBg24A}) with respect to $\lambda$ yields
\begin{equation}
\label{FRGBg24AB} \lambda = - \frac{2}{\mu^4} \left(
\frac{R}{2\kappa^2} + \frac{1}{2} T_\mathrm{matter}
 - 2 \tilde V \left( \chi \right) - 4 \left( - R^{\tau\eta}
+ \frac{1}{2} g^{\tau\eta} R \right) \nabla_\tau \nabla_\eta h
\left( \chi \right) \right) \, .
\end{equation}
The spatially flat Friedmann-Robertson-Walker (FRW) metric ansatz will fulfill
our purpose in the present context, in particular,
\begin{equation}
\label{FRWmetric} 
ds^2 = - dt^2 + a(t)^2 \sum_{i=1,2,3} \left( dx^i \right)^2 \, .
\end{equation}
Considering the functions $\lambda$ and $\chi$ depend only on cosmic time, and also that no matter fluids are present, that
is, that $T_{\mathrm{matter}\, \mu\nu} =0$, then Eq.~(\ref{FRGBg20}) allows the following simple solution
\begin{equation}
\label{frgdS4} \chi = \mu^2 t \, .
\end{equation}
Hence, the $(t,t)$ and $(i,j)$ components of Eq.~(\ref{FRGBg24}) can be written as
\begin{align}
\label{FRGFRW1} 
0 = & - \frac{3H^2}{2\kappa^2}
 - \frac{\mu^4 \lambda}{2} + \frac{1}{2} \tilde V \left( \mu^2 t \right)
 - 12 \mu^2 H^3 h' \left( \mu^2 t \right) \, , \\
\label{FRGFRW2} 
0 = & \frac{1}{2\kappa^2} \left( 2 \dot H + 3 H^2 \right)
 - \frac{1}{2} \tilde V \left( \mu^2 t \right)
+ 4 \mu^4 H^2 h'' \left( \mu^2 t \right) + 8 \mu^2 \left( \dot H +
H^2 \right) H h' \left( \mu^2 t \right) \, ,
\end{align}
and, in addition, from Eq.~(\ref{FRGBg23}) we get
\begin{equation}
\label{FRGFRW3} 
0 = \mu^2 \dot\lambda + 3 \mu^2 H \lambda + 24 H^2
\left( \dot H + H^2 \right) h'\left( \mu^2 t \right)
 - {\tilde V}'\left( \mu^2 t \right) \, .
\end{equation}
Eq.~(\ref{FRGFRW1}) is an algebraic equation with respect to $\lambda$, and thus we obtain,
\begin{equation}
\label{FRGFRW4} 
\lambda = - \frac{3 H^2}{\mu^4 \kappa^2} 
+ \frac{1}{\mu^4} \tilde V \left( \mu^2 t \right) - \frac{24}{\mu^2}
H^3 h' \left( \mu^2 t \right) \, .
\end{equation}
It is easy to see that, by combining Eqs.~(\ref{FRGFRW4}) and
(\ref{FRGFRW3}), we obtain Eq.~(\ref{FRGFRW2}). 
Moreover Eq.~(\ref{FRGFRW2}) leads to the scalar potential 
$\tilde V \left( \mu^2 t \right)$ as,
\begin{equation}
\label{FRGFRW7} \tilde V \left( \mu^2 t \right) =
\frac{1}{\kappa^2} \left( 2 \dot H + 3 H^2 \right) + 8 \mu^4 H^2
h'' \left( \mu^2 t \right) + 16 \mu^2 \left( \dot H + H^2 \right)
H h' \left( \mu^2 t \right) \, .
\end{equation}
Hence, for an arbitrarily chosen function $h(\chi (t))$, the potential $\tilde V \left( \chi \right)$ being equal to
\begin{equation}
\label{FRGFRW8} 
\tilde V \left( \chi \right) = \left[
\frac{1}{\kappa^2} \left( 2 \dot H + 3 H^2 \right) + 8 \mu^4 H^2
h'' \left( \mu^2 t \right) + 16 \mu^2 \left( \dot H + H^2 \right)
H h' \left( \mu^2 t \right) \right]_{t=\frac{\chi}{\mu^2}}\, ,
\end{equation}
then we can realize an arbitrary cosmology corresponding to a given Hubble rate $H(t)$. 
Finally, the functional form of the Lagrange multiplier reads
\begin{equation}
\label{FRGFRW4B} 
\lambda = \frac{2 \dot H}{\mu^4 \kappa^2} + 8 H^2
h'' \left( \mu^2 t \right) + \frac{8}{\mu^2} \left( 2 \dot H - H^2
\right) H h' \left( \mu^2 t \right) \, .
\end{equation}
As mentioned in the introductory section, here we are interested
on a smooth unified scenario from an ekpyrotic bounce to the dark energy era followed from the theory with the action 
(\ref{FRGBg22}). 
The resulting theory with the action (\ref{FRGBg22}) resembles with the scalar coupled Einstein-Gauss-Bonnet theory, in which case, the 
speed of gravitational wave ($c_T^2$) differs from unity and the deviation of $c_T^2$ from unity depends on the Gauss-Bonnet coupling function. In particular, 
the speed of gravitational wave in the present context comes with the following form 
\cite{Hwang:2005hb,Noh:2001ia,Hwang:2002fp}: 
\begin{equation}
c_T^2 = 1 + \frac{16\left( \ddot{h}-\dot{h}H \right)}{\frac{1}{\kappa^2} + 16\dot{h}H}
\label{gravitaional wave speed}
\end{equation}
with $H = \frac{\dot{a}}{a}$ being the Hubble parameter.
Eq.~(\ref{gravitaional wave speed}) apparently reflects the non-viability of the model with respect to GW170817 which validates the fact that
the gravitational and electromagnetic waves have the same propagation speed (i.e., unity in natural units). 
However the gravitational wave speed in the ghost free $f(R,\mathcal{G})$ model becomes $c_T^2 = 1$ 
if the coupling function satisfies the following constraint equation \cite{Odintsov:2020sqy,Odintsov:2020zkl},
\begin{eqnarray}
\ddot{h} = \dot{h}H\, .
\label{constraint on coupling}
\end{eqnarray}
The above constraint makes a ghost free $f(R,\mathcal{G})$ model consistent with the event GW170817. 
Thereby we need to consider such Gauss-Bonnet coupling 
functions which obey Eq.~(\ref{constraint on coupling}) in order to have a compatibility of the present model with GW170817. 
A form of the Hubble parameter fixes the Gauss-Bonnet coupling function by Eq.~(\ref{constraint on coupling}), 
by plugging which in Eq.~(\ref{FRGFRW8}), one gets the scalar potential. 
In the next section, we will consider a suitable Hubble parameter for our present purpose, and will reconstruct the form of 
$h(\chi)$ as well as of $V(\chi)$. 
At this stage it is worth mentioning that the constraint Eq.~(\ref{constraint on coupling}) on $h(\chi)$ also fits with the original equations of
motion,  this being clear from the fact that there are two 
independent equations, namely the $(t,t)$ component of the
gravitational equation and the equation for $\chi$, and yet two
unknown functions ($\lambda$(t), $V(\chi)$) to determine.

\section{Background evolution}\label{sec-background}

We are interested in unifying an ekpyrotic non-singular bounce with the present dark energy era, and for this purpose, the background 
scale factor is considered as \cite{Odintsov:2016tar},
\begin{eqnarray}
a(t) = \left[1 + a_0\left(\frac{t}{t_0}\right)^2\right]^{\frac{1}{3(1+w)}}
\exp{\left[\frac{1}{(\alpha-1)}\left( \frac{t_s - t}{t_0} \right)^{1-\alpha}\right]}\, ,
\label{scale factor1}
\end{eqnarray}
where $w$, $\alpha$, and $t_s$ are various parameters. $t_0$ is a fiducial time taken to make the above expression dimensionally 
correct, and we take $t_0 = 1\,\mathrm{By}$ (By stands for Billion years) in the subsequent calculation. 
We also assume $t<t_s$. 
In effect, the scale factor can be re-written as.
\begin{eqnarray}
a(t) = \left[1 + a_0t^2\right]^n\times \exp \left[\frac{1}{(\alpha-1)}\left( t_s - t \right)^{1-\alpha}\right] = a_1(t)\times a_2(t)\ (\mathrm{say})\, ,
\label{scale factor2}
\end{eqnarray}
with $n = \frac{1}{3(1+w)}$. 
It may be observed that $a(t)$ gets a product of $a_1(t)$ and $a_2(t)$, respectively. 
Actually $a_1(t)$, with $w>1$ or equivalently $n < 1/6$, is sufficient to get a non-singular ekpyrotic bounce, 
however, at large positive time, $a_1(t)$ behaves as $a_1(t) \sim t^{2n}$ which is not consistent with a viable dark energy era. 
Thus in order to unify an ekpyrotic bounce with a viable dark energy era, the scale factor is taken to be a product of $a_1(t)$ and $a_2(t)$, with $w>1$. 
Due to the exponential behaviour, the term $a_2(t)$ has almost no role in the contracting stage of the universe 
and thus the bouncing character is controlled by $a_1(t)$, except the fact that the presence of $a_2(t)$ slightly modifies the time of bounce. 
On contrary, $a_2(t)$ shows a significant contribution during the expanding phase of the universe, due to which, $a_2(t)$ 
along with $a_1(t)$ leads to a viable dark energy era in the current universe. 
In particular, the scale factor of Eq.~(\ref{scale factor2}) seems to smoothly 
unify an ekpyrotic bounce to the dark energy era with an intermediate deceleration era in-between the bounce and late time acceleration, and further, the 
theoretical expectations of various primordial as well as dark energy observables are found to be compatible with the Planck data for suitable 
parameter values. We will come to this point in details at some stage.

The Hubble parameter and the Ricci scalar, from Eq.~(\ref{scale factor2}), are determined as,
\begin{eqnarray}
H(t) = \frac{1}{a}\frac{da}{dt} = \frac{2a_0nt}{\left(1 + a_0t^2\right)} + \frac{1}{\left( t_s - t \right)^{\alpha}}
\label{Hubble parameter}
\end{eqnarray}
and
\begin{eqnarray}
R(t) = \frac{12a_0n}{\left(1 + a_0t^2\right)^2}\left\{1 - a_0t^2\left(1-4n\right)\right\} 
+ \frac{12}{\left( t_s - t \right)^{2\alpha}} + \frac{6\alpha}{\left( t_s - t \right)^{1+\alpha}} + \frac{48a_0nt}{\left(1 + a_0t^2\right)
\left( t_s - t \right)^{\alpha}} \, , 
\label{ricci scalar}
\end{eqnarray}
respectively. 
It is evident from Eq.~(\ref{Hubble parameter}) that depending on the value of $\alpha$, 
the Hubble parameter and/or the higher derivatives of the Hubble parameter diverge at $t = t_s$. 
In particular, the condition $\alpha > 1$ leads to a 
Type-I singularity where the scale factor, the effective energy density and the effective pressure simultaneously diverge at $t = t_s$, while 
for $\alpha < 1$, the model predicts a Type-II or Type-III or Type-IV singularity at $t = t_s$, depending on whether 
$\alpha > 0$ or $\alpha < 0$ (see \cite{Nojiri:2005sx} for finite time future singularity). 
Thus a finite time singularity at $t = t_s$ seems to be inevitable in the present context, irrespective of the values of $\alpha$. 
Therefore in order to describe a singularity free evolution of the universe 
up-to the present epoch $t_p \approx 13.5\,\mathrm{By}$ ($t_p$ stands for present time), we consider $t_s > t_p$. 
Here we may argue that in the future 
(i.e., at $t > t_p$) -- either the universe will face a finite time singularity predicted by the present model or possibly more fundamental 
theory will govern that regime by which the aforementioned singularity can be avoided. 
However due to the consideration $t_s > t_p \approx 13.5\,\mathrm{By}$, the model provides 
a singular free evolution of the universe at-least up-to the present epoch.

\begin{figure}[!h]
\begin{center}
\centering
\includegraphics[scale=0.70]{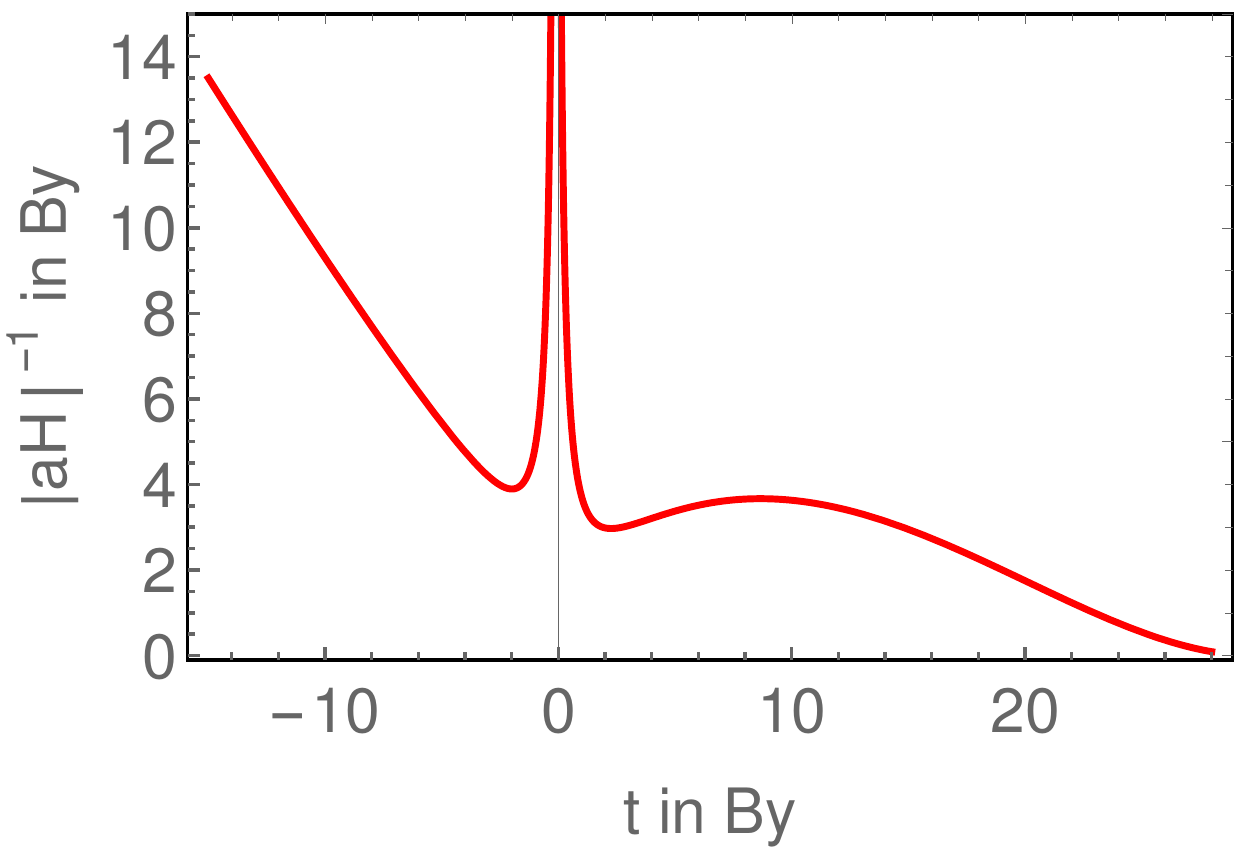}
\caption{$r_h$ vs. $t$ for $\alpha = 4/3$, $w = 3$, $t_s = 30\,\mathrm{By}$ and $a_0 = 0.35$.}
\label{plot-Hubble-radius}
\end{center}
\end{figure}

 From Eq.~(\ref{Hubble parameter}), we determine the comoving Hubble radius (defined by $r_h = 1/|aH|$) which is shown in the Fig.~[\ref{plot-Hubble-radius}], 
for a suitable set of parameter values that indeed lead to the compatibility 
between the theoretical expectations of various observables with their respective Planck data (as we will show later). 
Such evolution of $r_h$ is important to analyze the generation era of primordial perturbation modes. 
Fig.~[\ref{plot-Hubble-radius}] clearly demonstrates that the comoving Hubble radius monotonically increases at large negative time and consequently 
diverges at $t \rightarrow -\infty$: this indicates that the primordial perturbations generate at large negative time when all the perturbatin modes 
are in the deep sub-Hubble region. 
On other side, $r_h$ seems to monotonically decrease with time at large positive time, which points towards a late time acceleration era of the universe. 
Furthermore, we may observe that $r_h(t)$ is not symmetric about $t \approx 0$, i.e., $r_h(-t) \neq r_h(t)$. 
This is, however, expected, because the presence of the term $a_2(t)$ in Eq.~(\ref{scale factor2}) makes the scale factor asymmetric 
about $t \approx 0$, due to which, the comoving Hubble radius gets such an asymmetric behaviour.

By using the above form of Hubble parameter, one can reconstruct the Gauss-Bonnet (GB) coupling function as well as the scalar potential. 
Integrating Eq.~(\ref{constraint on coupling}) with respect to cosmic time yields,
\begin{eqnarray}
\dot{h}(\chi(t)) = \frac{(2n+1)h_0}{a_0^n}a(t) \quad \mbox{or} \quad 
h(\chi) = \frac{(2n+1)h_0}{a_0^n} \left. \int^{t}a(t)dt \right|_{t = \chi/\mu^2}\, ,
\label{reconstruct-1}
\end{eqnarray}
where $h_0$ is an integration constant having mass dimension $[2n+1]$. 
By plugging the expression of $a(t)$ in the above equation, we will obtain the explicit form of $\dot{h}(t)$ or $h(\chi)$ as well. 
However, the Friedmann equations as well as the scalar field equation of motion 
contain the derivative of GB coupling rather than the $h(\chi)$ itself, which is a consequence of the fact that a constant 
$h$ (essentially no coupling) would immediately make the GB contribution trivial. 
Thus it is not necessary to explicitly perform the integration in the expression of $h(\chi)$ in Eq.~(\ref{reconstruct-1}). 
Furthermore, owing to the constraint Eq.~(\ref{constraint on coupling}), the scalar potential 
and the Lagrange multiplier (from Eqs.~(\ref{FRGFRW8}) and (\ref{FRGFRW4B})) can be simplified and given by,
\begin{align}
V(\chi)=&\left(2\dot{H} + 3H^2\right)\left. \left(\frac{1}{\kappa^2} + 8\dot{h}H\right) \right|_{t=\chi/\mu^2}\, ,\nonumber\\
\lambda(t)=&\frac{2\dot{H}}{\mu^4}\left(\frac{1}{\kappa^2} - 8\dot{h}H\right)\, ,
\label{reconstruct-2}
\end{align}
respectively. 
In accordance of Eq.~(\ref{reconstruct-1}), the scalar potential and the Lagrange multiplier can be expressed as,
\begin{align}
V(\chi) =&\left(2\dot{H} + 3H^2\right) \left. \left(\frac{1}{\kappa^2} + \frac{8(2n+1)h_0}{a_0^n}a(t)H(t)\right) \right|_{t=\chi/\mu^2}\, ,\nonumber\\
\lambda(t) =&\frac{2\dot{H}}{\mu^4}\left(\frac{1}{\kappa^2} - \frac{8(2n+1)h_0}{a_0^n}a(t)H(t)\right)\, .
\label{reconstruct-3}
\end{align}
The above equation provides the $V(\chi)$ and $\lambda(t)$ in terms of the background scale factor and Hubble parameter. 
Therefore plugging the aforementioned expressions of $a(t)$ and $H(t)$ into Eq.~(\ref{reconstruct-3}), 
will immediately lead to the explicit form of $V(\chi)$ and $\lambda(t)$, respectively. 
We will use Eq.~(\ref{reconstruct-1}) and Eq.~(\ref{reconstruct-3}) in the subsequent calculations.

\subsection{A non-singular ekpyrotic bounce}\label{sec-bounce}
In this section, our aim is to examine whether the scale factor of Eq.~(\ref{scale factor2}) leads to a non-singular ekpyrotic bounce. 
For this purpose, we borrow the expression of the Hubble parameter from Eq.~(\ref{Hubble parameter}) as,
\begin{eqnarray}
H(t) = \frac{2a_0nt}{\left(1 + a_0t^2\right)} + \frac{1}{\left( t_s - t \right)^{\alpha}}\, .
\label{Hubble parameter-bounce1}
\end{eqnarray}
At the time of bounce, the universe makes a transition from a contracting stage to an expanding one, and thus, the Hubble parameter and its first derivative 
satisfy $H_b = 0$ and $\dot{H}_b > 0$, respectively (where the suffix 'b' stands for bounce). 
For $t > 0$, both the terms present in the right hand side of Eq.~(\ref{Hubble parameter-bounce1}) are positive; 
while for $t < 0$, the first term gets negative and the second term remains positive. 
Therefore we hope to have $H = 0$ and consequently a bounce at $t < 0$. In particular, for $t < 0$, we take $t = -|t|$ and 
Eq.~(\ref{Hubble parameter-bounce1}) is re-written as,
\begin{eqnarray}
H(t<0) = -\frac{2a_0n|t|}{\left(1 + a_0|t|^2\right)} + \frac{1}{\left(t_s + |t|\right)^{\alpha}} = -H_1(t) + H_2(t)~(\mathrm{say})\, .
\label{Hubble parameter-bounce2}
\end{eqnarray}
The functional forms of $H_1(t)$ and $H_2(t)$ clearly argue that both $H_1(t)$ and $H_2(t)$ tend to zero at $t \rightarrow -\infty$. 
On the other side, in particular at $t = 0^{-}$, $H_1(t) = 0$, while $H_2(t) = 1/t_s^{\alpha}$. During the interval $-\infty < t < 0$, $H_1(t)$ gets a maximum 
at $t = -1/\sqrt{a_0}$, while $H_2(t)$ monotonically increases in the aforementioned time interval. 
Moreover at $t \rightarrow -\infty$, the rate of increasing of $H_1(t)$ and $H_2(t)$ are determined as
\begin{eqnarray}
\left. \frac{dH_1}{dt}\right|_{t\rightarrow -\infty} \sim 1/|t|^2 \, , \quad \left. \frac{dH_2}{dt} \right|_{t\rightarrow -\infty} = 1/|t|^{(1+\alpha)}\, .\nonumber
\end{eqnarray}
The above expression indicates that for $\alpha > 1$ (which is indeed viable in respect to the Planck data and also a requirement to get a positive Ricci scalar 
during the expanding universe, as we will show later),, $H_1(t)$ increases at a faster rate in comparison to that of $H_2(t)$. 
Thus as a whole -- both $H_1(t)$ and $H_2(t)$ have 
an increasing behaviour with respect to time and start from the value zero at $t \rightarrow -\infty$, 
however $H_1(t)$ again reaches to zero at $t = 0^{-}$ with getting a maximum at $t = -1/\sqrt{a_0}$, while $H_2(t)$ reaches to a positive value 
at $t = 0^{-}$ with a monotonic increasing behaviour during $-\infty < t < 0$. 
Furthermore at $t \rightarrow -\infty$, the rate of increasing of $H_1(t)$ (with respect to time) seems to be larger compared to that of $H_2(t)$. 
Such informations ensure that at some point of time during $-\infty < t < 0$, $H_1(t)$ gets equal to $H_2(t)$ 
or equivalently $H = 0$, let the time being is $t = -\tau$. 
Depending on whether $t < -\tau$ or $t \geq -\tau$, the following statements about the Hubble parameter hold true: 
(1) $H_1(t) > H_2(t)$ or equivalently $H(t) < 0$ during $t < -\tau$, (2) $H_1(t) = H_2(t)$ 
or $H(t) = 0$ at $t = -\tau$ and (3) $H_1(t) < H_2(t)$ or $H(t) > 0$ during $t > -\tau$. 
Therefore $t = -\tau$ is the time instant when the bounce occurs, which can be determined from the condition $H(-\tau) = 0$, i.e., 
\begin{eqnarray}
\frac{2a_0n\tau}{\left(1 + a_0\tau^2\right)} = \frac{1}{\left(t_s + \tau\right)^{\alpha}}\, .
\label{bounce-instant}
\end{eqnarray}
Eq.~(\ref{bounce-instant}) is an algebraic equation for $\tau$, which may not be solved in a closed form. 
However for a clear picture, we estimate $\tau$ for a suitable set of parameter values. 
In particular, for $w = 3$, $a_0 = 0.35$, $\alpha = 4/3$ and $t_s = 30\,\mathrm{By}$ (which are viable in regard to the 
Planck data, as we will show later), we get $\tau = -0.13\,\mathrm{By}$. 
The scale factor at $\tau = -0.13\,\mathrm{By}$ becomes 
$a(\tau) = 2.624$, i.e., $a(t)$ is positive at the time of bounce, which indicates that the bounce is indeed non-singular. 
Here it may be mentioned that if the scale factor is controlled by $a_1(t)$ only, i.e., 
$a(t) = a_1(t)$ (the expression of $a_1(t)$ is given in Eq.~(\ref{scale factor2})), 
the time of bounce becomes $\tau = 0$ and moreover the scale factor gets symmetric around $\tau = 0$, 
unlike to the present context where the scale factor is given by $a(t) = a_1(t)\times a_2(t)$ (see Eq.~(\ref{scale factor2})) which leads to 
an asymmetric bounce and the bounce occurs at a negative $t$. 
Actually the presence of $a_2(t)$ slightly modifies the instant of bounce and also breaks the symmetric nature of the bounce, 
compared to the case where the $a_2(t)$ is absent. 

Coming back to Eq.~(\ref{Hubble parameter-bounce1}), the Hubble parameter at large negative time behaves as $H(t) = 2n/t$. 
Consequently the effective equation of state (EoS) parameter turns out to be,
\begin{eqnarray}
w_\mathrm{eff} = -1-\frac{2\dot{H}}{3H^2} = -1+\frac{1}{3n} = w\, ,
\label{eos-1}
\end{eqnarray}
where in the last equality, we use $n = \frac{1}{3(1+w)}$. 
Therefore in order to have an ekpyrotic nature of the bounce, the parameter $w$ is constrained 
to be $w > 1$. With the condition $w>1$, the bouncer field (the effective energy density coming from the Gauss-Bonnet higher curvature degrees of freedom 
acts as a bouncer in the present context) in the contracting universe decays more faster than the anisotropic energy density. 
In effect of which, the background evolution of the contracting stage remains stable to the growth of anisotropies and thus gets free from the BKL instability. 
Thus as a whole, the scale factor of Eq.~(\ref{scale factor2}) with $w > 1$ is able to describe a non-singular ekpyrotic bouncing universe. 

In regard to the evolution of the Ricci scalar, we start with the expression of $R(t)$ from Eq.~(\ref{ricci scalar}), i.e., 
\begin{eqnarray}
R(t) = \frac{12a_0n}{\left(1 + a_0t^2\right)^2}\left\{1 - a_0t^2\left(1-4n\right)\right\} 
+ \frac{12}{\left( t_s - t \right)^{2\alpha}} + \frac{6\alpha}{\left( t_s - t \right)^{1+\alpha}} + \frac{48a_0nt}{\left(1 + a_0t^2\right)
\left( t_s - t \right)^{\alpha}}\, .
\label{ricci scalar1}
\end{eqnarray}
It is clear that at $t \rightarrow -\infty$, the Ricci scalar behaves as $R(t) \approx -\frac{12n(1-4n)}{t^2}$, i.e., $R(t)$ is negative during the 
late contracting era. 
However, at the instant of bounce, the Hubble parameter vanishes and its first derivative becomes positive; in effect of which, 
$R(t) (=12H^2 + 6\dot{H})$ acquires positive value at the bounce. 
The above arguments indicate that the Ricci scalar experiences a zero crossing from negative to positive values before the bounce occurs. 
In such situation, we require that after the zero crossing, the Ricci scalar remains 
to be positive throughout the cosmic expansion of the universe, in particular, $R(t)$ should satisfy,
\begin{eqnarray}
R(t > -\tau) > 0\, ,
\label{requirement1}
\end{eqnarray}
where recall, $t = -\tau$ is the time when the bounce happens. 
The constraint in Eq.~(\ref{requirement1}) lead to the following conditions between the model parameters:
\begin{eqnarray}
\alpha > 1 \quad \mbox{and} \quad 
\sqrt{a_0} > \frac{4\sqrt{3}}{t_s^{\alpha}\left(1-4n\right)} \, ,
\label{model parameter constraint-1}
\end{eqnarray}
respectively. 
These conditions are obtained by the following demonstrations: it is evident that the Ricci scalar diverges at $t = t_s$. 
Therefore in order to examine Eq.~(\ref{requirement1}), we need to investigate the behaviour of $R(t)$ 
during $-\tau < t < t_s$. in this regime, $R(t)$ can be expressed as,
\begin{eqnarray}
R(t) = \frac{12a_0n}{\left(1 + a_0t^2\right)^2}\left\{1 + \frac{4}{t_s^{\alpha}}t - a_0t^2\left(1 - 4n\right) + \frac{4a_0}{t_s^{\alpha}}t^3\right\} 
+ \frac{12}{t_s^{2\alpha}} + \frac{6\alpha}{t_s^{1+\alpha}}\, .
\label{ricci scalar2}
\end{eqnarray}
The only term, due to which the Ricci scalar may acquire negative values during the expanding phase, is given by $a_0t^2(1-4n)$ present within the 
curly bracket of Eq.~(\ref{ricci scalar2}), and moreover the last two terms in the above expression contribute positive values to $R(t)$. 
Thereby to examine Eq.~(\ref{requirement1}), we can safely consider the terms that are present within the curly bracket in the right hand side of 
Eq.~(\ref{ricci scalar2}), let us denote it by $\widetilde{R}(t)$, in particular,
\begin{eqnarray}
\widetilde{R}(t) = \left\{1 + \frac{4}{t_s^{\alpha}}t - a_0t^2\left(1 - 4n\right) + \frac{4a_0}{t_s^{\alpha}}t^3 \right\}\, .
\label{tilde-R}
\end{eqnarray}
Consequently, Eq.~(\ref{requirement1}) can be equivalently written as 
\begin{eqnarray}
\widetilde{R}(t > -\tau) > 0\, .
\label{requirement-sub}
\end{eqnarray}
It can be shown from Eq.~(\ref{tilde-R}) that $\widetilde{R}(t)$ possesses a maximum and a minimum at,
\begin{eqnarray}
t_\mathrm{max} = \frac{a_0\left(1-4n\right) - \sqrt{a_0^2(1-4n)^2 - 48a_0/t_s^{\alpha}}}{12a_0/t_s^{\alpha}}
\label{maximum}
\end{eqnarray}
and
\begin{eqnarray}
t_\mathrm{min} = \frac{a_0\left(1-4n\right) + \sqrt{a_0^2(1-4n)^2 - 48a_0/t_s^{\alpha}}}{12a_0/t_s^{\alpha}} \, ,
\label{minimum}
\end{eqnarray}
respectively. 
Accordingly, we have,
\begin{eqnarray}
\widetilde{R}(t_\mathrm{max}) = 1 + \frac{2t_\mathrm{max}\left\{1 - a_0~t_\mathrm{max}^2\right\}}{t_s^{\alpha}}\, ,\nonumber\\
\widetilde{R}(t_\mathrm{min}) = 1 + \frac{2t_\mathrm{min}\left\{1 - a_0~t_\mathrm{min}^2\right\}}{t_s^{\alpha}}\, .
\label{function value}
\end{eqnarray}
It is clear that $\widetilde{R}(t_\mathrm{min}) > 0$ (i.e., the minimum value of $\widetilde{R}(t)$ is positive) in turn leads to the condition 
$\widetilde{R}(t > -\tau) > 0$ or equivalently $R(t>-\tau) > 0$. 
Now Eq.~(\ref{function value}) indicates that the minimum of $\widetilde{R}(t)$, 
in particular, $\widetilde{R}(t_\mathrm{min})$ gets positive if the model parameters obey Eq.~(\ref{model parameter constraint-1}).

Thus as a whole, the Ricci scalar starts from $R(t) \rightarrow 0^{-}$ at $t \rightarrow -\infty$, and experiences a zero crossing 
from negative to positive values before the instant of bounce. 
However after the zero crossing, the Ricci scalar is found to be positive during the expansion of the universe once 
the model parameters follow the condition in Eq.~(\ref{model parameter constraint-1}). 
Therefore owing to the requirement $R(t>-\tau) > 0$, we stick to the condition of Eq.~(\ref{model parameter constraint-1}) in the present work.

\subsection{Deceleration and acceleration stages of the universe}\label{sec-acc-deceleration}
The acceleration factor of the universe is given by $\frac{\ddot{a}}{a} = \dot{H} + H^2$ which, in the present context, turns out to be,
\begin{eqnarray}
\frac{\ddot{a}}{a} = \frac{2a_0n\left\{1 - a_0t^2(1 - 2n)\right\}}{\left(1 + a_0t^2\right)^2} 
+ \frac{\alpha}{\left( t_s - t \right)^{1+\alpha}} + \frac{4a_0nt}{\left(1 + a_0t^2\right)\left( t_s - t \right)^{\alpha}} 
+ \frac{1}{\left( t_s - t \right)^{2\alpha}}\, .
\label{acceleration-1}
\end{eqnarray}
Before moving to the quantitative description of $\ddot{a}/a$, first we qualitatively analyze the same for different phases of the universe, 
in particular, for the contracting and expanding phases, respectively.

\subsubsection*{{\underline{During the contraction}}}
At late contracting era, i.e., at $t \rightarrow -\infty$, the acceleration factor can be approximated by,
\begin{eqnarray}
\frac{\ddot{a}}{a} \approx -\frac{2n(1-2n)}{t^2}\label{contraction-1}\, ,
\end{eqnarray}
where recall, $n = \frac{1}{3(1+w)}$ with $w$ being the effective EoS parameter at large negative time. 
In order to have an ekpyrotic contraction phase, $w$ should be greater than unity. 
As a result, Eq.~(\ref{contraction-1}) clearly indicates that the universe undergoes through a deceleration phase at $t \rightarrow -\infty$. 
However as the universe approaches towards the bounce, the term $a_0t^2(1-2n)$ (present within the curly bracket 
in the right hand side of Eq.~(\ref{acceleration-1})) gets smaller than unity and consequently $\ddot{a}$ becomes positive. 
In particular, the acceleration factor near the bounce can be expressed as,
\begin{eqnarray}
\frac{\ddot{a}}{a} \approx 2a_0n + \frac{\alpha}{t_s^{1+\alpha}} + \frac{1}{t_s^{2\alpha}}\, ,
\label{contraction-2}
\end{eqnarray}
which indeed indicates an accelerating universe. 
This is, however, expected, because the bounce regime is, by definition, an accelerating phase of the universe. 
Thereby as a whole, the universe in the contracting era starts from a decelerating phase at $t \rightarrow -\infty$, and during the evolution, it transits 
from the deceleration phase to an accelerating one. 
Such transition from deceleration to acceleration occurs nearly when the term $a_0t^2(1-2n)$ becomes unity, 
in particular the time of transition (say, at $t_1$) can be determined from,
\begin{eqnarray}
a_0~t_1^2(1-2n) = 1 \quad \Longrightarrow \quad t_1 = -\frac{1}{\sqrt{a_0\left(1-2n\right)}}\label{contractoion-3}\, .
\end{eqnarray}

\subsubsection*{{\underline{During the expansion}}}
During the expanding phase, the only term that contributes negative value to $\ddot{a}$ is given by $a_0t^2(1-2n)$ present within the curly bracket 
in the right hand side of Eq.~(\ref{acceleration-1}). thereby due to the competition between $a_0t^2(1-2n)$ and the 
other terms of Eq.~(\ref{acceleration-1}), the universe in the expanding stage experiences several transitions from acceleration to deceleration or vice-versa. 
The demonstration goes as follows: as $t$ increase from zero, the term $a_0t^2(1-2n)$ starts to grow. 
In particular, during $a_0t^2(1-2n) > 1$, the first term of Eq.~(\ref{acceleration-1}) becomes negative and hence the universe may expand through a decelerating stage. 
As $t$ further increases and approaches to $t = t_s$, the terms containing $1/\left(_s - t\right)$ grows at a faster rate compared to the other terms, 
and as a result, $\ddot{a}$ may become positive, i.e., the universe transits from the decelerating phase to an accelerating one. 
The transition from deceleration $\longrightarrow$ acceleration or vice-versa can be 
defined by $\ddot{a} = 0$ which, from Eq.~(\ref{acceleration-1}), is written as,
\begin{eqnarray}
\frac{2a_0n\left\{1 - a_0t^2(1 - 2n)\right\}}{\left(1 + a_0t^2\right)^2} 
+ \frac{\alpha}{\left( t_s - t \right)^{1+\alpha}} + \frac{4a_0nt}{\left(1 + a_0t^2\right)\left( t_s - t \right)^{\alpha}} 
+ \frac{1}{\left( t_s - t \right)^{2\alpha}} = 0\, .
\label{acceleration-2}
\end{eqnarray}
Eq.~(\ref{acceleration-2}) may not be solved in a closed form. 
However based on the above arguments, we determine the solutions of Eq.~(\ref{acceleration-2}) 
in different regimes, in particular, during $\frac{1}{\sqrt{a_0\left(1-2n\right)}} < t \ll t_s$ and 
$\frac{1}{\sqrt{a_0\left(1-2n\right)}} \ll t < t_s$, respectively.

\begin{enumerate}
\item During $\frac{1}{\sqrt{a_0\left(1-2n\right)}} < t \ll t_s$: In this regime, Eq.~(\ref{acceleration-2}) can be approximated by,
\begin{eqnarray}
\frac{4a_0n}{t_s^{\alpha}}t^3 - 2a_0n(1-2n)t^2 + 2n = 0\, ,
\label{expansion-1}
\end{eqnarray}
which has the solution near at,
\begin{eqnarray}
t_2 = \frac{1}{\sqrt{a_0\left(1-2n\right)}}\left(1 + \frac{a_0}{t_s^{\alpha}\left(a_0(1-2n)\right)^{3/2}}\right)\, .
\label{expansion-2}
\end{eqnarray}
Thus the universe during the expanding stage experiences its first transition from an acceleration 
(occurs near the bounce) to a deceleration at $t = t_2$.
\item During $\frac{1}{\sqrt{a_0\left(1-2n\right)}} \ll t < t_s$: In this regime, Eq.~(\ref{acceleration-2}) may be re-written as,
\begin{eqnarray}
\frac{1}{t_s^{2\alpha}}t^2 + \frac{2\alpha\left(1-2n\right)}{t_s}t - (1-2n) = 0\, .
\label{expansion-3}
\end{eqnarray}
The above algebraic equation has the solution at,
\begin{eqnarray}
t_3 = \alpha t_s^{2\alpha - 1}\left(1 - 2n\right)
\left\{ \sqrt{1 + \frac{1}{\alpha^2\left(1 - 2n\right)t_s^{2\alpha - 2}}} - 1 \right\}\, ,
\label{expansion-4}
\end{eqnarray}
where we consider the positive root. 
Thereby the universe during the expanding stage makes the second and final transition from the deceleration phase to 
an accelerating phase at $t = t_3 > t_2$.
\end{enumerate}

As a whole, in the expanding phase, the universe experiences two consecutive transitions -- (1) the first transition occurs 
at $t = t_2$ from acceleration (happens near the bounce) to a deceleration and (2) the second transition is at $t = t_3$ from the deceleration to 
a late time acceleration. 
In order to confront the model with supernovae results, the final acceleration phase is identified with the current dark energy epoch. 
For this purpose, we require $t_3 < t_p = 13.5\,\mathrm{By}$. 
 From Eq.~(\ref{expansion-4}), we can write $t_3 = t_s/2\alpha$ by binomially expanding the square root and retain up-to the leading order term 
(due to $\alpha > 1$). 
Therefore the inequality $t_3 < t_p$ can be equivalently expressed by 
\begin{eqnarray}
t_s < 2\alpha t_p\, .
\label{expansion-5}
\end{eqnarray}
Furthermore, as mentioned earlier, $t = t_s$ denotes a Type-I singularity and thus in order to describe a singular free universe up-to the present epoch, the parameter 
$t_s$ should be greater than $t_p$, i.e., $t_s > t_p$. 
Combining this with Eq.~(\ref{expansion-5}), we get the final constraint on $t_s$ as,
\begin{eqnarray}
t_p < t_s < 2\alpha t_p\, ,
\label{expansion-6}
\end{eqnarray}
which is indeed valid due to $\alpha > 1$.

\subsection{Constraints on model parameters}\label{sec-constraints}
In the present context, there are four model parameters in total: $n$ (or equivalently $w$ by the relation $n = \frac{1}{3(1+w)}$), $\alpha$, $a_0$ and $t_s$. 
In the following, we list the possible constraints on such model parameters, coming from different considerations. 
Some of the constraints are discussed in the previous subsections, however here we put all the constraints one by one to obtain a more clear picture.
\begin{itemize}
\item $\mathrm{C1}$: In order to have an ekpyrotic phase of contraction to avoid the BKL instability, the parameter $w$ should be 
greater than unity (or equivalently $n < 1/6$).
\item $\mathrm{C2}$: As mentioned earlier, $t = t_s$ denotes a Type-I singularity and thus to get a singular free description of the universe up-to the 
present epoch, $t_s$ should satisfy $t_s > t_p = 13.5\,\mathrm{By}$.
\item $\mathrm{C3}$: the final accelerating stage of the universe is identified with the current dark energy epoch. 
For this purpose, $t_s < 2\alpha t_p$ as discussed in Eq.~(\ref{expansion-5}). 
This, along with the previous condition (i.e., $\mathrm{C2}$), immediately leads  to $t_p < t_s < 2\alpha t_p$.
\item $\mathrm{C4}$: For the parameters $\alpha$ and $a_0$, they are constrained as 
$\alpha > 1$ and $\sqrt{a_0} > \frac{4\sqrt{3}}{t_s^{\alpha}\left(1-4n\right)}$, respectively. 
These make the Ricci scalar positive during the expanding phase of the universe. 
Actually by these constraints, the Ricci scalar remains positive throughout the cosmic evolution after its zero crossing 
from negative to positive values, where the zero crossing occurs before the bounce happens.
\item $\mathrm{C5}$: The effective equation of state (EoS) parameter is defined as $w_\mathrm{eff} = -1 - \frac{2\dot{H}}{3H^2}$, where $H(t)$ is shown 
in Eq.~(\ref{Hubble parameter}). 
In order to confront the dark energy epoch of the present model with Planck+SNe+BAO results, the effective 
EoS parameter at present epoch is constrained to be \cite{Aghanim:2018eyx},
\begin{eqnarray}
w_\mathrm{eff}(t_p) = -0.957 \pm 0.080\, ,
\end{eqnarray}
where recall, $t_p = 13.5\,\mathrm{By}$ represents the present cosmic time of the universe.
\item $\mathrm{C6}$: The Planck data further leads to a bound on the dark energy relic density as \cite{Aghanim:2018eyx},
\begin{eqnarray}
\Omega_\mathrm{D} = 0.6847 \pm 0.0073\, ,
\end{eqnarray}
which will be taken into account in the present context.
\end{itemize}

Keeping the above constraints in mind, we may take, for example, $w = 3$ and $\alpha = 4/3$. 
Consequently the condition $\mathrm{C3}$ leads to $13.5 < t_s < 36$ (in the unit of $\,\mathrm{By}$), and thus we safely take $t_s = 30\,\mathrm{By}$. 
With such values of $w$, $\alpha$ and $t_s$, the constraint $\mathrm{C4}$ leads to $a_0 > 0.012$, while $\mathrm{C5}$ and $\mathrm{C6}$ lead to 
$a_0 > 0.097$ and $0.184 < a_0 < 2.220$, respectively. 
Accounting all the above conditions, the constraint on $a_0$ turns out to be: $0.184 < a_0 < 2.220$. Based on this inequality, $a_0 = 0.35$ gives 
the theoretical expectation of relic dark energy density near its central value of the Planck result, in particular, 
$\Omega_\mathrm{D}(a_0 = 0.35) = 0.6848$. 
As a whole, if the model parameters are considered to be, $w = 3$ (or equivalently $n=\frac{1}{12}$), $\alpha = 4/3$, $a_0 = 0.35$ and 
$t_s = 30\,\mathrm{By}$; the theoretical expectations of the dark energy EoS parameter and the relic dark energy density are obtained as,
\begin{eqnarray}
w_\mathrm{eff}(t_p) = -0.961 \quad \mbox{and} \quad \Omega_\mathrm{D} = 0.6848\, ,
\nonumber
\end{eqnarray}
respectively, which are simultaneously compatible with the Planck results. 
Here it may be mentioned that in order to estimate the dark energy EoS parameter as well as the relic dark energy density, the following conversations 
may be useful: $1\mathrm{GeV} = 1.52\times10^{24}\mathrm{sec}^{-1}$ and $1\,\mathrm{By}^{-1} = \left(\frac{3.17}{1.52}\right)\times10^{-41}\mathrm{GeV}$, 
respectively. 

\begin{figure}[!h]
\centering
\includegraphics[scale=0.35]{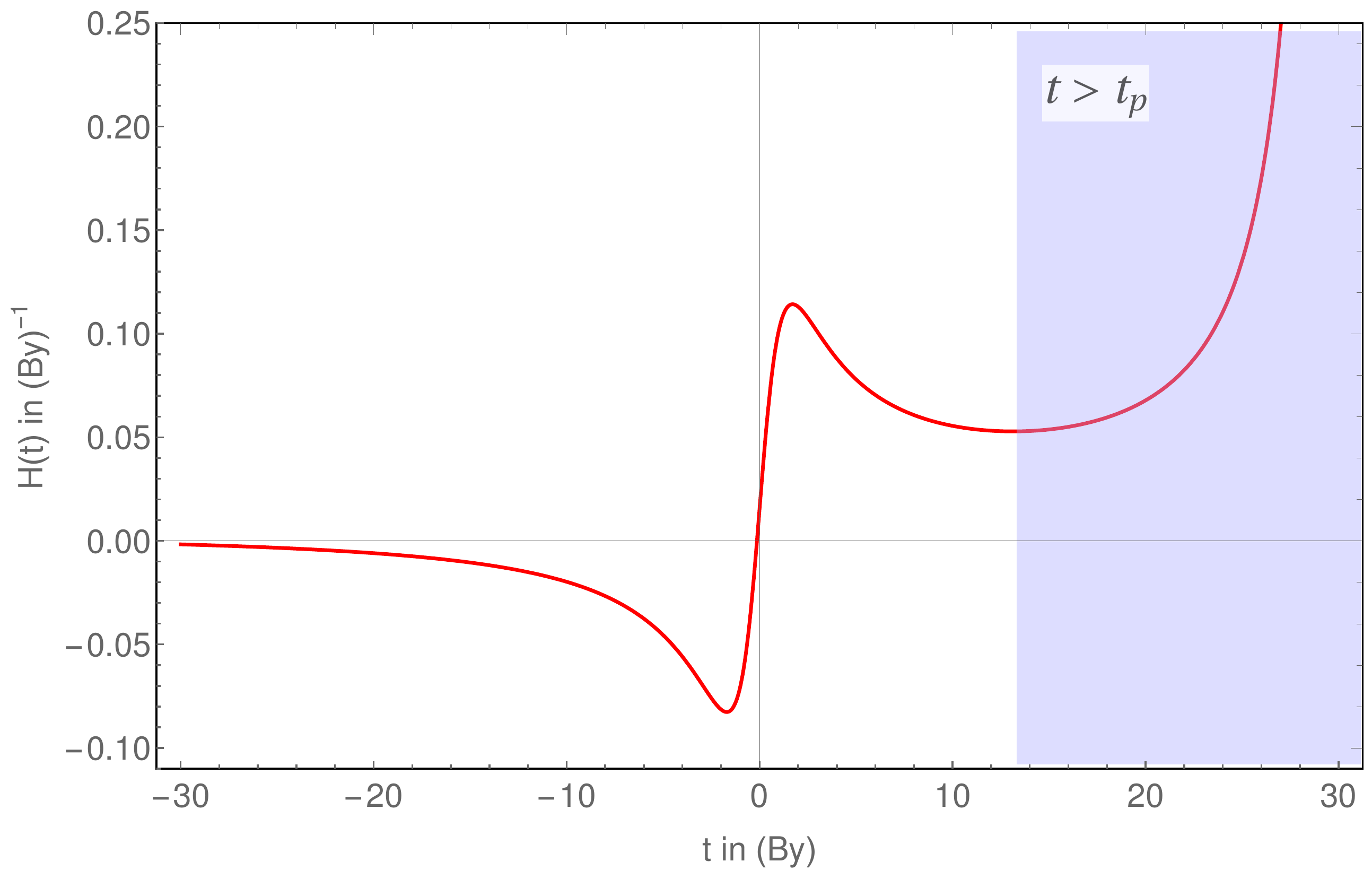}
\includegraphics[scale=0.32]{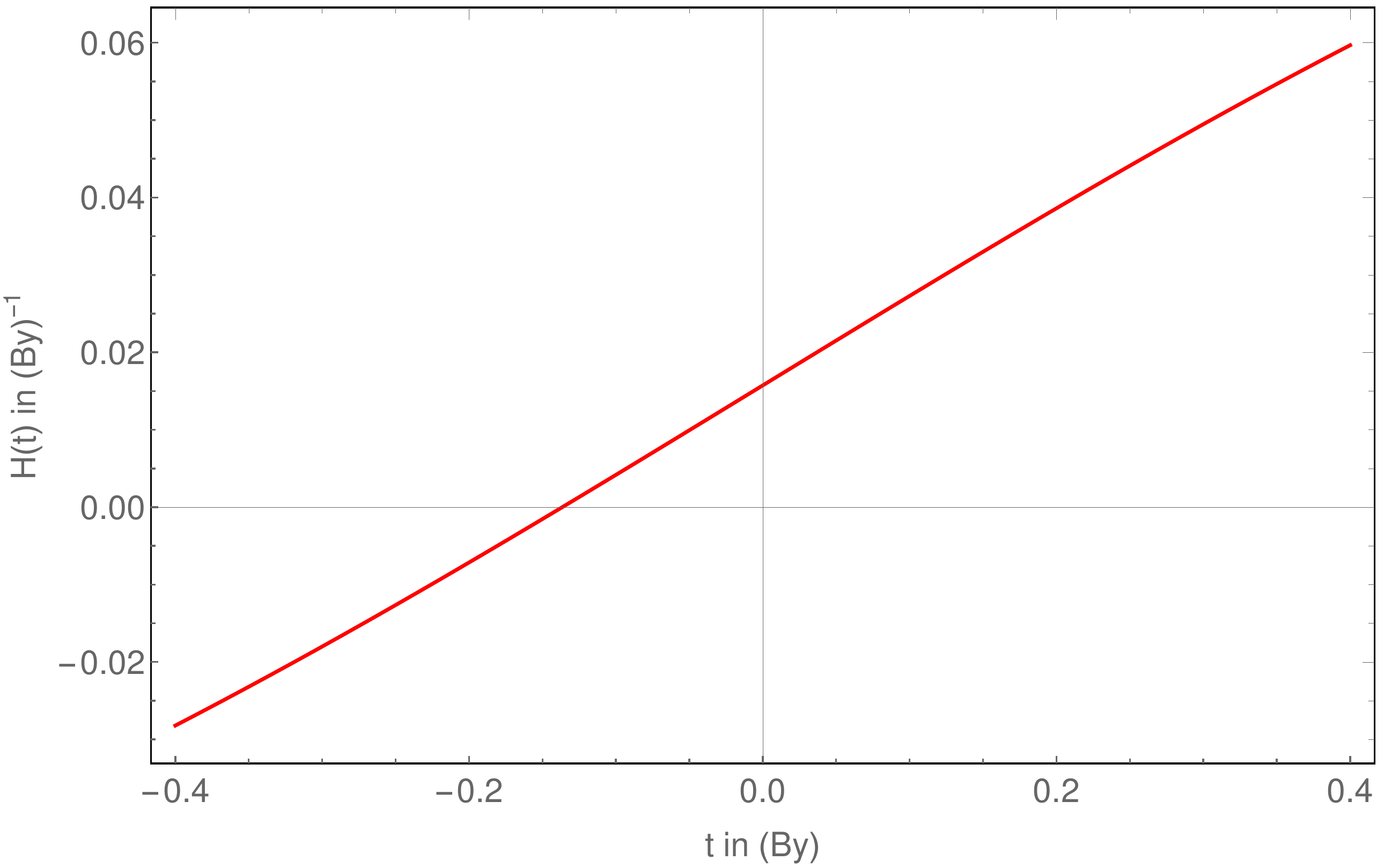} 
\caption{{\underline{Left Plot}}: $H(t)$ vs. $t$ for $w=3$, $\alpha = 4/3$, $a_0 = 0.35$ and $t_s = 30\,\mathrm{By}$, respectively. 
The shaded region represents the cosmic time with $t > t_p = 13.8\,\mathrm{By}$. 
{\underline{Right Plot}}: The zoomed-in version of $H(t)$ vs. $t$ near the bounce where the Hubble parameter vanishes. In both the plots, 
the $H(t)$ is in the unit of $\,\mathrm{By}^{-1}$ and $t$ is in $\,\mathrm{By}$}
\label{plot-Hubble-parameter}
\end{figure}

\begin{figure}[!h]
\centering
\includegraphics[scale=0.40]{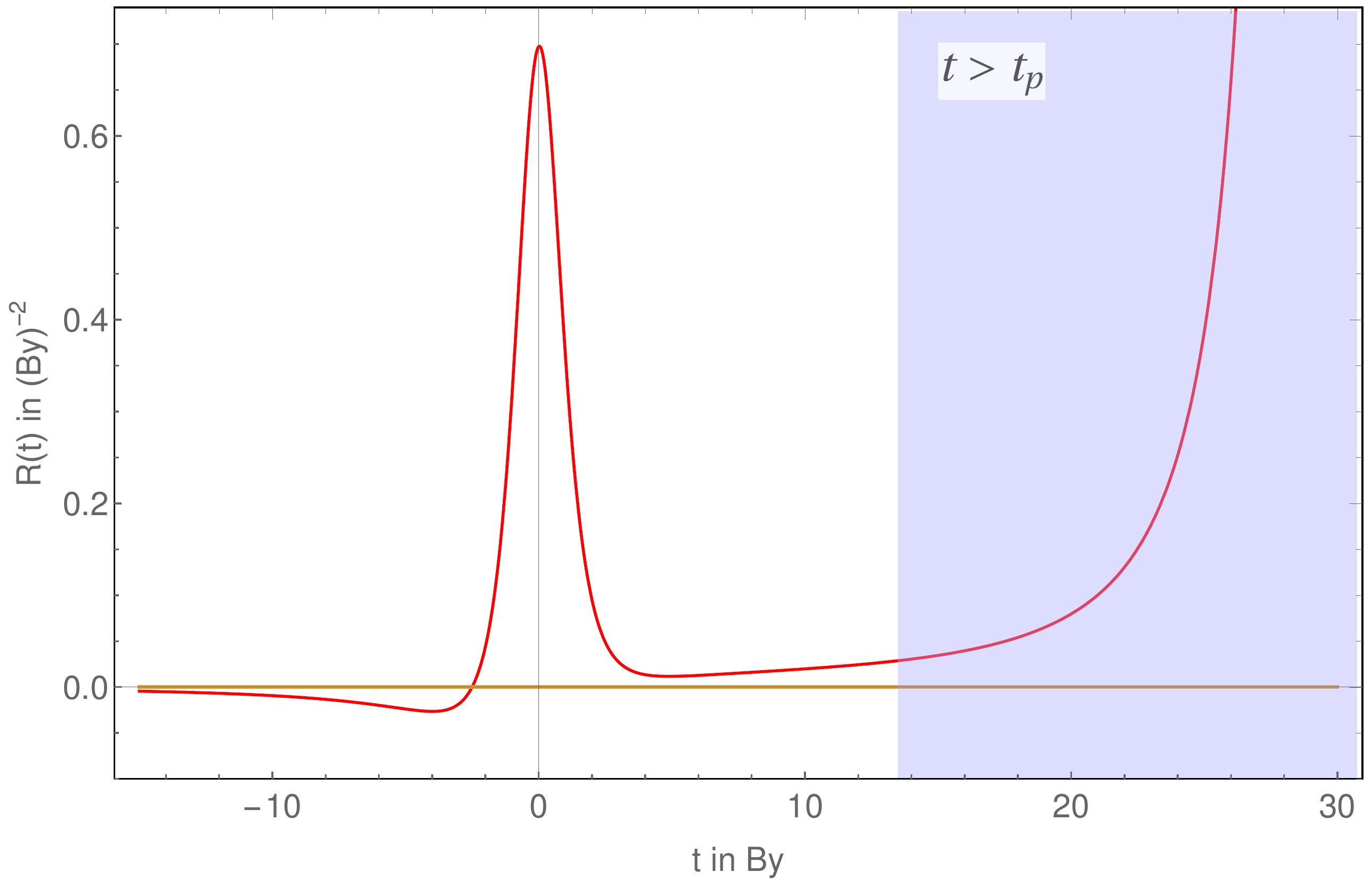}
\caption{$R(t)$ vs. $t$ for $w=3$, $\alpha = 4/3$, $a_0 = 0.35$ and $t_s = 30\,\mathrm{By}$, respectively. The $R(t)$ is in the unit 
of $\,\mathrm{By}^{-2}$ and $t$ is in $\,\mathrm{By}$. It is evident that the zero crossing of the Ricci scalar occurs before the bounce, and after the zero 
crossing, it remains positive throughout the cosmic time. Moreover the shaded region denotes $t>t_p = 13.8\,\mathrm{By}$.}
\label{plot-ricci-scalar}
\end{figure}

Taking all the constraints into account (i.e., from $\mathrm{C1}$ to $\mathrm{C6}$), 
we give the plots of the background Hubble parameter, Ricci scalar and the effective equation of state (EoS) parameter. 
In particular, we consider $w=3$, $\alpha = 4/3$, $a_0 = 0.35$ and $t_s = 30\,\mathrm{By}$ in the plots, which indeed lie within the aforementioned constraints. 
The left plot of Fig.~[\ref{plot-Hubble-parameter}] clearly demonstrates that the Hubble parameter vanishes and increases with time near $t \approx 0$, 
which thus refers to a non-singular bounce. 
Actually $t = -0.13\,\mathrm{By}$ indicates the time of bounce, see the right plot of Fig.~[\ref{plot-Hubble-parameter}] 
which is a zoomed-in version of the Hubble parameter near the bounce. 
This is in agreement with the discussion of Sec.[\ref{sec-bounce}] where we showed that due to the presence of the factor $a_2(t)$ 
in the scale factor, the instant of bounce is slightly shifted to a negative time. 
In regard to the evolution of the Ricci scalar, Fig.~[\ref{plot-ricci-scalar}] reveals that the $R(t)$ starts its journey from $0^{-}$ at $t\rightarrow -\infty$, 
however as the time increases, $R(t)$ experiences a zero crossing from negative to positive values. 
Here it may be mentioned that the zero crossing of $R(t)$ occurs before the bounce happens, which is expected, 
because at the bounce, the Ricci scalar ($R(t) = 12H^2 + 6\dot{H}$) acquires positive values. 
After such zero crossing, the $R(t)$ seems to be positive throughout the cosmic evolution of the universe. 
Actually the parameter values considered in the plot satisfy the constraint $\mathrm{C4}$ which in turn leads to the $positive$ values of 
the Ricci scalar during the expanding phase of the universe, as discussed in Sec.[\ref{sec-bounce}]. 
At this stage, it deserves mentioning that both the Hubble parameter and the Ricci scalar 
seem to diverge at $t = t_s$ (recall, $t_s = 30\,\mathrm{By}$ in the plots), which in turn refers to a Type-I singularity. 
However, the important point to be observed that $t=t_s$ lies far away from the present time, i.e., from $t=t_p = 13.8\,\mathrm{By}$. 
Thereby we may argue that the current model safely describes a singular free evolution of the universe up-to $t \gtrsim t_p$. 

\begin{figure}[!h]
\centering
\includegraphics[scale=0.55]{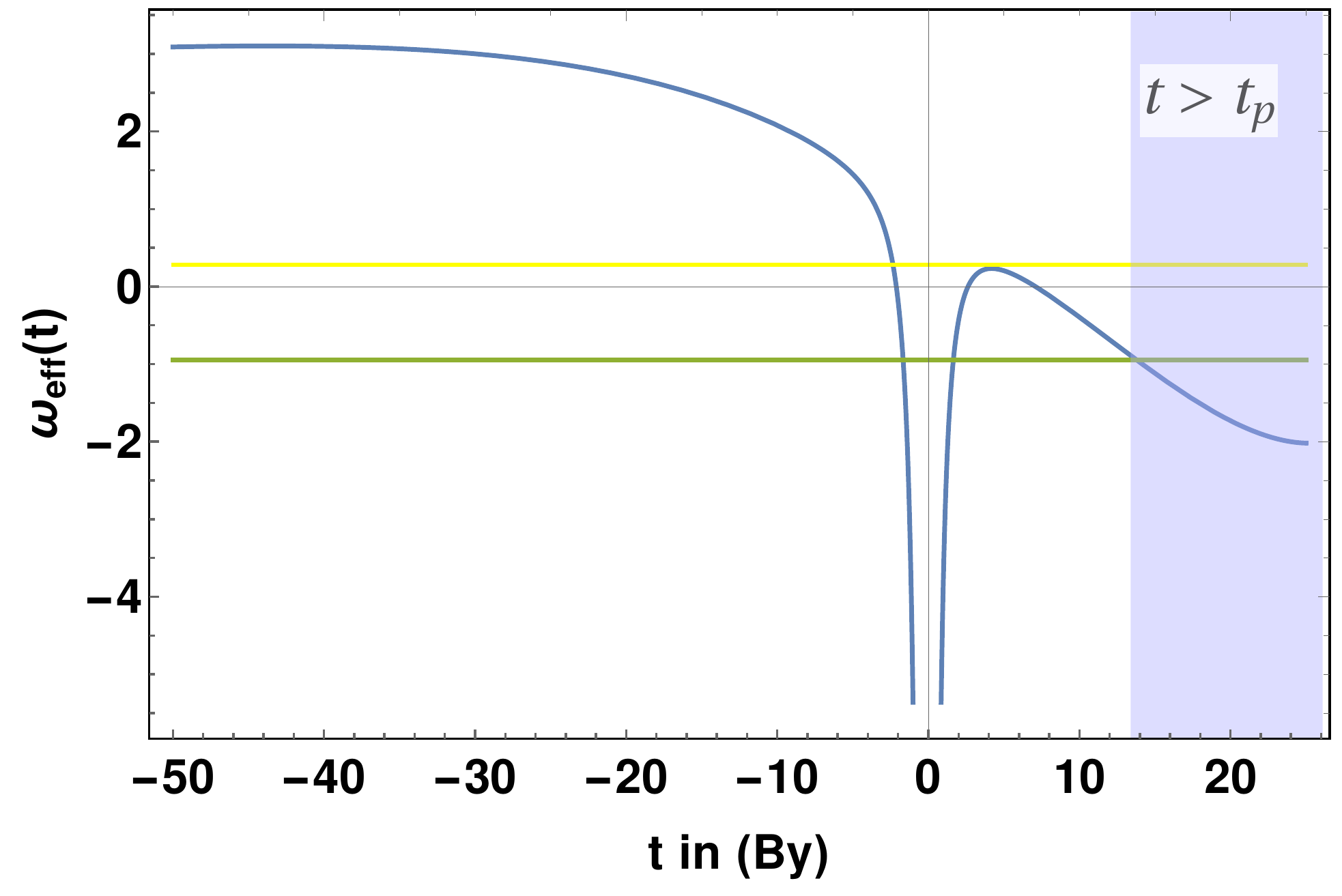}
\caption{$w_\mathrm{eff}(t)$ vs. $t$ for $w=3$, $\alpha = 4/3$, $a_0 = 0.35$ and $t_s = 30\,\mathrm{By}$, respectively. The 
left part of $t = 0$ deals with $w_\mathrm{eff}$ in the contracting phase of the universe. 
It seems that the effective EoS parameter during the contracting phase remains larger than unity, which ensures an ekpyrotic phase of contraction 
that is free from BKL instability. On other hand, the right of $t = 0$ shows $w_\mathrm{eff}(t)$ during 
the expanding phase of the universe. Moreover, the horizontal yellow and green curves represent 
the constant values $1/3$ and $-0.961$, respectively. The intersection between the green curve and the $w_\mathrm{eff}(t)$ clearly 
indicates that the effective EoS parameter acquires the value $=-0.961$ at present time.}
\label{plot-eos}
\end{figure}

In regard to the effective EoS parameter, we refer to the Fig.~[\ref{plot-eos}] which demonstrates the evolution of $w_\mathrm{eff}(t)$ with respect to $t$. 
The left part of $t = 0$ in the figure deals with $w_\mathrm{eff}(t)$ during the contracting phase of the universe, which depicts that 
$w_\mathrm{eff} \approx w = 3$ (recall, we take $w=3$ in the plot) at late contracting era. 
This is however expected, as the scale factor behaves as $a(t) \sim t^{\frac{2}{3(1+w)}}$ and consequently $w_\mathrm{eff} = w$ at large negative time. 
Furthermore, Fig.~[\ref{plot-eos}] shows that the effective EoS parameter during the contracting stage remains 
larger than unity, which ensures an ekpyrotic phase of contraction where the BKL instability can be avoided. 
Thereby the bounce universe in the present context undergoes through an ekpyrotic phase in the contracting stage and thus 
becomes free from the BKL instability.

The right part of $t = 0$ in Fig.~[\ref{plot-eos}] shows the $w_\mathrm{eff}(t)$ during the expanding phase, where 
the horizontal yellow curve represents the constant value $1/3$. At $t \rightarrow 0^{+}$, the EoS parameter diverges to $w_\mathrm{eff} \rightarrow -\infty$. 
This is however expected, because $t \approx 0$ is the bounce regime where the Hubble parameter vanishes 
and consequently $w_\mathrm{eff} = -1-\frac{2\dot{H}}{3H^2} \rightarrow$ tend to $-\infty$. As $t$ increases during the expanding phase, 
$w_\mathrm{eff}$ crosses the value $-\frac{1}{3}$ (from $w_\mathrm{eff} < -\frac{1}{3}$ to $w_\mathrm{eff}> -\frac{1}{3}$), which 
actually indicates a transition of the universe's evolution from an accelerating phase to a deceleration one. 
During the deceleration phase, the EoS parameter acquires a maximum value of $w_\mathrm{eff} = \frac{1}{3}$ -- that leads to a radiation-like era. Consequently, the EoS parameter 
during the deceleration stage passes through $w_\mathrm{eff} = 0$ resulted to a mater-like era. 
As $t$ further increases, $w_\mathrm{eff}$ again crosses the value $-\frac{1}{3}$ (from $w_\mathrm{eff} > -\frac{1}{3}$ to $w_\mathrm{eff} < -\frac{1}{3}$), i.e., the universe 
makes its final transition from the deceleration phase to an accelerating one. 
This accelerating stage is identified with the current dark energy epoch. 
The horizontal green curve in the plot represents the constant value $-0.961$ that seems to intersect with the curve of $w_\mathrm{eff}$ at $t=t_p=13.8\,\mathrm{By}$. 
Therefore, for the parameter values considered in the figure, the theoretical expectation of dark energy EoS parameter becomes $w_\mathrm{eff}(t_p) = -0.961$, 
which is indeed compatible with the Planck+SNe+BAO results \cite{Aghanim:2018eyx}. 
Actually, such values of the parameters satisfy the aforementioned constraint $\mathrm{C5}$ which in turn leads to the compatibility between the 
theoretical estimation and the Planck observational result of the dark energy EoS parameter.

\section{Cosmological perturbation and primordial observable quantities}\label{sec-perturbation}
In this section, we consider fluctuation over the FRW metric and consequently calculate various primordial 
observable quantities like scalar spectral index, tensor to scalar ratio etc. In bouncing universe, the primordial perturbation modes 
generate either near the bounce or far away from the bounce depending on the asymptotic behaviour of comoving Hubble radius (defined by $r_h = \left(aH\right)^{-1}$). 
For a bounce to happen, the Hubble parameter becomes zero and thus the Hubble radius diverges at the instant of bounce; however, 
the asymptotic evolution of Hubble radius at deep contracting era makes the bounce models different to each other. 
In some bounce scenario, the Hubble radius monotonically decreases and asymptotically goes to zero at the deep contracting era, in which case, the perturbation modes 
generate near the bounce, because at that time, the Hubble size becomes infinity and thus all the modes lie within the sub-Hubble domain. 
On other hand, the Hubble radius in some bounce model(s) increases and diverges to infinity at $t \rightarrow -\infty$, which in turn makes the 
generation era of perturbation modes far away from the bounce at the contracting era. 
Therefore before moving to the solution of perturbation variables, it is important to examine the generation era of perturbation modes in the present context where 
an ekpyrotic bounce gets smoothly unified to dark energy era in the backdrop of the ghost free Gauss-Bonnet gravity theory. 

As mentioned by Fig.~[\ref{plot-Hubble-radius}], the Hubble radius in the present context tends to infinity at $t \rightarrow -\infty$, 
and thus the perturbation modes generate far away from the bounce in the deep sub-Hubble regime. 
This is a direct consequence of the fact that the scale factor at large negative time behaves as $a(t) \sim t^{\frac{2}{3(1+w)}}$ with $w > 1$ due 
to an ekpyrotic phase of contraction. 
Here it may be mentioned that the present bounce scenario is an asymmetric one, in particular, unlike to the fact that $r_h$ diverges at 
$t \rightarrow -\infty$, it monotonically decreases at large positive time, which actually leads to the dark energy era of tge universe. 
Furthermore, at the deep contracting era, the background scale factor, Hubble parameter and the Gauss-Bonnet coupling function (in terms of the 
Ricci scalar) turn out to be,
\begin{align}
a(R) = a_0^{n}/\widetilde{R}^n\, ,& \quad H(R) = -2n\widetilde{R}^{1/2}\, ,\nonumber\\
\dot{H}(R) = -2n\widetilde{R}\, ,& \quad \dot{h}(R) = h_0(2n+1)/\widetilde{R}^n\, ,
\label{quantities-deep-contracting}
\end{align}
where $\widetilde{R}(t) = \frac{-R(t)}{12n(1-4n)} > 0$ and the overdot represents $\frac{d}{dt}$. 
Here we need to recall that the Ricci scalar at large negative time acquires negative values, in particular $R(t) \rightarrow 0^{-}$ at $t \rightarrow -\infty$ (see 
Fig.~[\ref{plot-ricci-scalar}]), which in turn makes the quantity $\widetilde{R}(t)$ positive valued. 
Using the above expressions of $H(R)$ and $\dot{h}(R)$, we determine the functions $Q_i$ in the context of 
the Lagrange multiplier Gauss-Bonnet theory of gravity \cite{Hwang:2005hb,Noh:2001ia,Hwang:2002fp} (which will be useful for the perturbation calculations) as,
\begin{align}
Q_a=&-8\dot{h}H^2 = -32h_0n^2(1+2n)\widetilde{R}^{1-n}\, , \nonumber\\
Q_b=&-16\dot{h}H = 32h_0n(1+2n)\widetilde{R}^{\frac{1}{2} - n}\, ,\nonumber\\
Q_c=&Q_d = 0 \, ,\nonumber\\
Q_e=&-32\dot{h}\dot{H} = 64h_0n(1+2n)\widetilde{R}^{1-n}\, ,\nonumber\\
Q_f=&16 \left[ \ddot{h} - \dot{h}H \right] = 0 \, ,
\label{Q-s}
\end{align}
respectively. 
Recall, $h_0$ has mass dimension [1+2n] and thus from dimensional analysis, we can consider,
\begin{eqnarray}
\kappa^2h_0 = \left(\frac{1}{R_0}\right)^{\frac{1}{2} - n}\, ,
\label{h0-dimension}
\end{eqnarray}
where $R_0$ is a positive constant and can be regarded as a replacement of $h_0$. 
Consequently Eq.~(\ref{reconstruct-3}) 
along with the Eq.~(\ref{quantities-deep-contracting}) immediately lead to the Lagrange multiplier function as,
\begin{eqnarray}
\mu^4\lambda = -\frac{4n\widetilde{R}}{\kappa^2}\left\{1 + 16n(1+2n)\left(\frac{\widetilde{R}(t)}{R_0}\right)^{\frac{1}{2} - n}\right\}\, .
\label{LM-deep-contracting}
\end{eqnarray}
We will use the these expressions in addressing the evolution of scalar and tensor perturbations in the following two subsections, respectively. 

\subsection{Scalar perturbations}
The scalar perturbation over FRW metric is defined as,
\begin{align}
ds^2 = -(1 + 2\Psi)dt^2 + a(t)^2(1 - 2\Psi)\delta_{ij}dx^{i}dx^{j}\, ,
\label{sp1}
\end{align}
where $\Psi(t,\vec{x})$ symbolizes the scalar perturbation variable. 
Here we work in the comoving gauge, in which case, the curvature perturbation ($\mathcal{R}(t,\vec{x})$) becomes identical 
with the above mentioned $\Psi(t,\vec{x})$ and thus we proceed with the variable $\Psi(t,\vec{x})$. 
The second order perturbed action for $\Psi(t,\vec{x})$ is given by \cite{Hwang:2005hb,Noh:2001ia,Hwang:2002fp},
\begin{align}
\delta S_{\psi} = \int dt d^3\vec{x} a(t) z(t)^2\left[\dot{\Psi}^2
 - \frac{c_s^2}{a^2}\left(\partial_i\Psi\right)^2\right]\, ,
\label{sp2}
\end{align}
where $z(t)$ and $c_s^2$ (the speed of the scalar perturbation wave), in the context of the Lagrange multiplier Gauss-Bonnet gravity, 
have the following expressions \cite{Hwang:2005hb},
\begin{align}
z(t) = \frac{a(t)}{H + \frac{Q_a}{2F + Q_b}} \sqrt{-\mu^4\lambda + \frac{3Q_a^2 + Q_aQ_e}{2F + Q_b}}
\label{sp3}
\end{align}
and
\begin{align}
c_{s}^{2} = 1 + \frac{Q_aQ_e/\left(2F + Q_b\right)}{-\mu^4\lambda + 3\frac{Q_a^2}{2F + Q_b}} \, ,
\label{speed scalar perturbation}
\end{align}
respectively, with $F=\frac{1}{2\kappa^2}$ in our case and the functions $Q_i$ are defined earlier. 
Eq.~(\ref{sp2}) depicts that the kinetic term for $\Psi(t,\vec{x})$ comes with positive sign (or equivalently the scalar perturbation becomes stable) for the condition $z(t)^2 > 0$. 
Later we will show that this condition indeed holds true in the present context, which in turn makes the scalar perturbation stable. 
Using Eq.~(\ref{quantities-deep-contracting}), we determine various terms present in the expression of $z(t)$ as follows,
\begin{align}
\frac{a(t)}{H + \frac{Q_a}{2F + Q_b}} = -\frac{a_0^{n}}{\widetilde{R}^{\frac{1}{2} + n}}
\left[\frac{1}{2n\left\{1 + \frac{16n(1+2n)\left(\widetilde{R}/R_0\right)^{\frac{1}{2} - n}}
{1 + 32n(1+2n)\left(\widetilde{R}/R_0\right)^{\frac{1}{2} - n}}\right\}}\right]
\nonumber
\end{align}
and
\begin{eqnarray}
&-&\mu^4\lambda + \frac{3Q_a^2}{2F + Q_b} + \frac{Q_aQ_e}{2F + Q_b}\nonumber\\
&=&\frac{4n\widetilde{R}}{\kappa^2}\left[1 + 16n(1+2n)\left(\widetilde{R}/R_0\right)^{\frac{1}{2} - n} 
+ \frac{768n^3(1+2n)^2\left(\widetilde{R}/R_0\right)^{1-2n}}{1 + 32n(1+2n)\left(\widetilde{R}/R_0\right)^{\frac{1}{2} - n}}
 - \frac{512n^2(1+2n)^2\left(\widetilde{R}/R_0\right)^{1-2n}}{1 + 32n(1+2n)\left(\widetilde{R}/R_0\right)^{\frac{1}{2} - n}}\right]\, ,
\nonumber
\end{eqnarray}
respectively. 
Plugging back the above expressions into Eq.~(\ref{sp3}) yields,
\begin{align}
z(t) = -\frac{a_0^n}{\kappa \widetilde{R}^n}~\frac{\sqrt{P(R)}}{Q(R)}
\label{sp4}
\end{align}
where $P(R)$ and $Q(R)$ are given by,
\begin{align}
P(R) = 4n\left[1 + 16n(1+2n)\left(\widetilde{R}/R_0\right)^{\frac{1}{2} - n} 
+ \frac{768n^3(1+2n)^2\left(\widetilde{R}/R_0\right)^{1-2n}}{1 + 32n(1+2n)\left(\widetilde{R}/R_0\right)^{\frac{1}{2} - n}}
 - \frac{512n^2(1+2n)^2\left(\widetilde{R}/R_0\right)^{1-2n}}{1 + 32n(1+2n)\left(\widetilde{R}/R_0\right)^{\frac{1}{2} - n}}\right]\, ,
\label{P}
\end{align}
and
\begin{align}
Q(R) = 2n\left[1 + \frac{16n(1+2n)\left(\widetilde{R}/R_0\right)^{\frac{1}{2} - n}}
{1 + 32n(1+2n)\left(\widetilde{R}/R_0\right)^{\frac{1}{2} - n}}\right]\, ,
\label{Q}
\end{align}
respectively. 
As mentioned earlier, the perturbation modes generate far away from the bounce, where 
the Ricci scalar satisfies $\frac{\widetilde{R}}{R_0} < 1$ as $R \rightarrow 0$ at $t \rightarrow -\infty$ 
(the demonstration of such condition is given after Eq.~(\ref{hc-1})). 
Thereby Eq.~(\ref{sp4}) clearly indicates that $z^2(t)$ is indeed positive and thus ensures the stability of scalar perturbation in the present context. 
For the purpose of perturbation evolution, it will be more useful if we switch ourselves to conformal time defined by $\eta = \int\frac{dt}{a(t)}$. 
Now the scale factor of Eq.~(\ref{scale factor2}) behaves as $a(t) \sim t^{2n}$ (recall, $n = \frac{1}{3(1+w)}$) at distant past and thus the conformal time is given by,
\begin{eqnarray}
\eta(t) = \left[\frac{1}{a_0^n(1-2n)}\right]t^{1-2n}\, .
\label{conformal time}
\end{eqnarray}
The parameter $w$ is constrained by $w > 1$ (see $C1$) or equivalently $n < 1/6$, by which, $\eta(t)$ seems to be a monotonic 
increasing function of cosmic time. Eq.~(\ref{conformal time}) immediately leads to the Ricci scalar in terms of $\eta$ as,
\begin{eqnarray}
\widetilde{R}(\eta) = \frac{1}{\left[a_0^n(1-2n)\right]^{2/(1-2n)}}\times\frac{1}{\eta^{2/(1-2n)}} \propto \frac{1}{\eta^{2/(1-2n)}}\, .
\label{ricci-scalar-conformal-time}
\end{eqnarray}
Plugging back the above expression into Eq.~(\ref{sp4}) yields the factor $z$ in terms of the conformal time as,
\begin{eqnarray}
z(\eta) \propto \left(\frac{\sqrt{P(\eta)}}{Q(\eta)}\right)\eta^{2n/(1-2n)}\, ,
\label{z-eta}
\end{eqnarray}
where $P(\eta) = P(R(\eta))$ and $Q(\eta) = Q(R(\eta))$. 
Accordingly we determine $\frac{1}{z}\frac{d^2z}{d\eta^2}$ which is essential for the scalar Mukhanov-Sasaki equation,
\begin{eqnarray}
\frac{1}{z}\frac{d^2z}{d\eta^2} = \frac{\xi(\xi - 1)}{\eta^2}\left\{1 - \frac{16n(1-4n^2)}{1-4n}\left(\frac{\widetilde{R}}{R_0}\right)^{\frac{1}{2} - n} 
+ \mathcal{O}\left(\frac{\widetilde{R}}{R_0}\right)^{1-2n}\right\}
\label{derivative-z-eta}
\end{eqnarray}
with $\xi = \frac{2n}{(1-2n)}$ and we use $\frac{d\widetilde{R}}{d\eta} = \frac{-2}{(1-2n)}\frac{\widetilde{R}}{\eta}$. 
Furthermore the speed of the scalar perturbation from Eq.~(\ref{speed scalar perturbation}) is obtained as,
\begin{eqnarray}
c_s^2 = 1 + \mathcal{O}\left(\frac{\widetilde{R}}{R_0}\right)^{1-2n}\, .
\label{sound-speed}
\end{eqnarray}
Having set the stage, we now introduce the scalar Mukhanov-Sasaki (MS) variable as $v(\eta,\vec{x}) = z(\eta)\mathcal{R}(\eta,\vec{x})$ 
which satisfies the following equation (known as Mukhanov-Sasaki equation), 
\begin{eqnarray}
\frac{d^2v_k(\eta)}{d\eta^2} + \left(c_s^2k^2 - \frac{1}{z}\frac{d^2z}{d\eta^2}\right)v_k(\eta) = 0\, ,
\label{scalar-MS-equation}
\end{eqnarray}
where $v_k(\eta)$ is the scalar MS variable in Fourier space for $k$-th mode. 
Clearly the evolution of $v_k(\eta)$ depends on the background quantity $z''/z$ (the prime denotes $\frac{d}{d\eta}$), 
and as a result, the MS equation may not be solved in a closed form. 
However in the present context, Eq.~(\ref{scalar-MS-equation}) has analytic solutions at large negative time, as we now show. 
Owing to the condition $\frac{\widetilde{R}}{R_0} < 1$ (as mentioned after Eq.~(\ref{Q})), the expressions of $z''/z$ and $c_s^2$ 
(from Eq.~(\ref{derivative-z-eta}) and Eq.~(\ref{sound-speed})) can be approximated as,
\begin{eqnarray}
\frac{1}{z}\frac{d^2z}{d\eta^2} \approx \frac{\xi(\xi - 1)}{\eta^2}\left\{1 - \frac{16n(1-4n^2)}{1-4n}
\left(\frac{\widetilde{R}}{R_0}\right)^{\frac{1}{2} - n}\right\} \quad \mbox{and} \quad c_s^2 \approx 1\, ,
\label{approximate-behaviour}
\end{eqnarray}
respectively, by retaining the terms up-to the order $\left(\widetilde{R}/R_0\right)^{\frac{1}{2} - n}$. 
Furthermore $\frac{\widetilde{R}}{R_0} \ll 1$ along with the constraint $n < 1/6$ (or equivalently $w > 1$ in order to have an ekpyrotic phase of contraction) demonstrate that 
the term $\left(\widetilde{R}/R_0\right)^{\frac{1}{2} - n}$ within the paranthesis can be safely considered to be small during the late contracting era. 
As a result, $z''/z$ becomes proportional to $1/\eta^2$, i.e., $\frac{1}{z}\frac{d^2z}{d\eta^2} = \sigma/\eta^2$, with,
\begin{eqnarray}
\sigma = \xi(\xi - 1)\left\{1 - \frac{16n(1-4n^2)}{1-4n}\left(\frac{\widetilde{R}}{R_0}\right)^{\frac{1}{2} - n}\right\}\, ,
\label{sigma}
\end{eqnarray}
which is approximately a constant during the era when the perturbation modes generate deep inside the Hubble radius. In effect along with 
$c_s^2 = 1$, we solve the scalar MS variable from Eq.~(\ref{scalar-MS-equation}) and given by,
\begin{eqnarray}
v(k,\eta) = \frac{\sqrt{\pi|\eta|}}{2}\left[c_1(k)H_{\nu}^{(1)}(k|\eta|) + c_2(k)H_{\nu}^{(2)}(k|\eta|)\right]\, ,
\label{scalar-MS-solution}
\end{eqnarray}
with $\nu = \sqrt{\sigma + \frac{1}{4}}$. 
Moreover $H_{\nu}^{(1)}(k|\eta|)$ and $H_{\nu}^{(2)}(k|\eta|)$ are the Hermite functions (having order 
$\nu$) of first and second kind, respectively, and $c_1$, $c_2$ are integration constants. 
We consider the Bunch-Davies state as initial condition, in particular, $\lim_{k|\eta| \gg 1}v(k,\eta) = \frac{1}{\sqrt{2k}}e^{-ik\eta}$. 
The fact that the perturbation modes generate in the deep sub-Hubble radius ensures the viability of the 
Bunch-Davies initial condition in the present context. 
Owing to the Bunch-Davies condition, the integration constants turn out to be $c_1 = 0$ and $c_2 = 1$, respectively. 
Consequently, the scalar power spectrum for $k$th mode comes with the following expression,
\begin{eqnarray}
\mathcal{P}_{\Psi}(k,\eta) = \frac{k^3}{2\pi^2} \left| \frac{v(k,\eta)}{z(\eta)} \right|^2 
= \frac{k^3}{2\pi^2} \left| \frac{\sqrt{\pi|\eta|}}{2z (\eta)}H_{\nu}^{(2)}(k|\eta|) \right|^2\, ,
\label{scalar-power-spectrum}
\end{eqnarray}
where in the second equality, we use the solution of $v(k,\eta)$. 
The horizon crossing condition for $k$th mode is given by $k = |aH|$ which, due to the expressions of scale factor and Hubble parameter 
in Eq.~(\ref{quantities-deep-contracting}), takes the following form,
\begin{eqnarray}
k = \frac{1}{\left| \eta_h\right|}\left(\frac{2n}{1-2n}\right) \quad \Rightarrow \quad k\left| \eta_h\right| = \frac{2n}{1-2n}\, ,
\label{hc-1}
\end{eqnarray}
where the suffix 'h' represents the horizon crossing instant. 
Eq.~(\ref{hc-1}) immediately leads to the horizon crossing instant for large scale modes, 
in particular for $k = 0.002\mathrm{Mpc}^{-1}$ (around which we will determine the observable quantities), as 
$\eta_h \approx -10\,\mathrm{By}$ where we consider $n = \frac{1}{12}$ (or equivalently $w = 3$). 
Therefore the large scale modes cross the horizon at $\approx -10\,\mathrm{By}$, and consequently the Ricci scalar 
at the horizon crossing of large scale modes becomes $|R| \sim 10^{-3}\,\mathrm{By}^{-2}$. 
Thereby the ratio $\left|\frac{R}{R_0}\right| = 10^{-3}$ with $R_0 = 1\,\mathrm{By}^{-2}$, which justifies our consideration 
$\left|\frac{R}{R_0}\right| \ll 1$ (and retain the terms up-to the leading order 
of $\left|\frac{R}{R_0}\right|$ in the expression of $z(t)$) for the purpose of determining the observable quantities. 
Here it may be mentioned that $R_0 = 1\,\mathrm{By}^{-2}$ is indeed viable with the Planck constraints in regard to the observable quantities, as we will show later. 
Furthermore, Eq.~(\ref{hc-1}) depicts the sub-Hubble and super-Hubble regime of $k$-th mode as,
\begin{align}
k\left|\eta\right|>\frac{2n}{1-2n}\, :& \, \mathrm{sub~Hubble~regime}\, ,\nonumber\\
k\left|\eta\right|<\frac{2n}{1-2n}\, :& \, \mathrm{super~Hubble~regime}\, .
\label{hc-2}
\end{align}
The parameter $n$ is constrained to be $n < 1/6$ (or equivalently $w > 1$) from the requirement of an ekpyrotic phase of contraction, 
by which, the quantity $\frac{2n}{1-2n}$ becomes less than unity. 
Therefore from Eq.~(\ref{hc-2}), the super-Hubble regime can be equivalently expressed by $k\left|\eta\right| \ll 1$. 
In effect, the scalar power spectrum (from Eq.~(\ref{scalar-power-spectrum})) in the super-Hubble scale turns out to be,
\begin{eqnarray}
\mathcal{P}_{\Psi}(k,\eta) = \left[\left(\frac{1}{2\pi}\right)\frac{1}{z\left|\eta\right|}\frac{\Gamma(\nu)}{\Gamma(3/2)}\right]^2
\left(\frac{k|\eta|}{2}\right)^{3-2\nu}\, .
\label{scalar-power-spectrum-superhorizon}
\end{eqnarray}
By using Eq.~(\ref{scalar-power-spectrum-superhorizon}), we can determine the spectral index of the primordial curvature perturbations (symbolized 
by $n_s$). However before proceeding to calculate $n_s$, we will determine the tensor power spectrum, which is necessary for 
evaluating the tensor to scalar ratio.

\subsection{Tensor perturbation}
The tensor perturbation over FRW metric is expressed as,
\begin{align}
ds^2 = -dt^2 + a(t)^2\left(\delta_{ij} + h_{ij}\right)dx^idx^j\, ,
\label{tp1}
\end{align}
where $h_{ij}(t,\vec{x})$ is the tensor perturbation variable which is indeed a gauge invariant quantity. 
The tensor perturbed action (up-to quadratic order) comes as \cite{Hwang:2005hb,Noh:2001ia,Hwang:2002fp},
\begin{align}
\delta S_{h} = \int dt d^3\vec{x} a(t) z_T(t)^2\left[\dot{h_{ij}}\dot{h^{ij}}
 - \frac{1}{a^2}\left(\partial_kh_{ij}\right)^2\right]\, ,
\label{tp2}
\end{align}
where $z_T(t)$, in the context of the Lagrange multiplier Gauss-Bonnet gravity theory, is \cite{Hwang:2005hb}
\begin{align}
z_T(t) = a\sqrt{F + \frac{1}{2}Q_b}\, .
\label{tp3}
\end{align}
Here $F = \frac{1}{2\kappa^2}$ and the function $Q_b$ is shown in Eq.~(\ref{Q-s}). 
We need to recall that the Gauss-Bonnet coupling function obeys $\ddot{h} = \dot{h}H$, by which, the speed of gravitational wave 
in the present context becomes unity and the model gets compatible with the event GW170817. 
Using the expression of $a(R)$ from Eq.~(\ref{quantities-deep-contracting}), we determine $z_T$ as, 
\begin{align}
z_T=&\frac{a_0^n}{\sqrt{2}\kappa\widetilde{R}^n}\left\{1 + 32n(1+2n)\left(\frac{\widetilde{R}}{R_0}\right)^{\frac{1}{2} - n}\right\}^{1/2}\nonumber\\
=&\frac{a_0^n}{\sqrt{2}\kappa\widetilde{R}^n}\left\{1 + 16n(1+2n)\left(\frac{\widetilde{R}}{R_0}\right)^{\frac{1}{2} - n} 
+ \mathcal{O}\left(\frac{\widetilde{R}}{R_0}\right)^{1-2n}\right\}\, ,
\label{zT-1}
\end{align}
where in the second line, we use the binomial expansion due to $\frac{\widetilde{R}}{R_0} < 1$. 
Eq.~(\ref{zT-1}) clearly indicates that $z_T^2(t) > 0$ which makes the kinetic term of $h_{ij}$ positive and consequently the tensor perturbation becomes stable. 
In order to express $z_T$ in terms of the conformal time ($\eta$), one needs the functional form of $\widetilde{R} = \widetilde{R}(\eta)$. 
Thus, according to $\widetilde{R}(\eta) \propto \eta^{-2/(1-2n)}$, the quantity $z_T(\eta)$ becomes,
\begin{eqnarray}
z_T(\eta) \propto S(R(\eta))\eta^{2n/(1-2n)}\, ,
\label{zT-2}
\end{eqnarray}
with $S(R(\eta))$ has the following expression,
\begin{eqnarray}
S(R(\eta)) = 1 + 16n(1+2n)\left(\frac{\widetilde{R}}{R_0}\right)^{\frac{1}{2} - n} + \mathcal{O}\left(\frac{\widetilde{R}}{R_0}\right)^{1-2n}\, .
\label{S}
\end{eqnarray}
Consequently, we determine $z_T''/z_T$ as follows,
\begin{eqnarray}
\frac{1}{z_T}\frac{d^2z_T}{d\eta^2} = \frac{\xi(\xi-1)}{\eta^2}\left\{1 + \frac{32n(1-4n^2)}{1-4n}\left(\frac{\widetilde{R}}{R_0}\right)^{\frac{1}{2} - n} 
+ \mathcal{O}\left(\frac{\widetilde{R}}{R_0}\right)^{1-2n}\right\}\, ,
\label{derivative-zT}
\end{eqnarray}
we use $\frac{d\widetilde{R}}{d\eta} = \frac{-2}{(1-2n)}\frac{\widetilde{R}}{\eta}$ and recall, $\xi = \frac{2n}{1-2n}$. 
The above expression will be useful for solving the tensor Mukhanov-Sasaki equation. 
As mentioned earlier, the perturbation modes in the present scenario generate 
as well as cross the horizon at the deep contracting era where the condition $\frac{\widetilde{R}}{R_0} < 1$ holds as 
$R \rightarrow 0$ at $t \rightarrow -\infty$. 
In effect of which, $z_T''/z_T$ becomes proportional to $1/\eta^2$, i.e., $\frac{1}{z_T}\frac{d^2z_T}{d\eta^2} = \sigma_T/\eta^2$, with
\begin{eqnarray}
\sigma_T = \xi(\xi - 1)\left\{1 + \frac{32n(1-4n^2)}{1-4n}\left(\frac{\widetilde{R}}{R_0}\right)^{\frac{1}{2} - n}\right\}\, , 
\label{sigma-T}
\end{eqnarray}
where we retain the leading order term of $\widetilde{R}/R_0$. 
Having set the stage, we now intend to solve the tensor Mukhanov-Sasaki (MS) equation given by,
\begin{align}
\frac{d^2v_T(k,\eta)}{d\eta^2} + \left(k^2 - \frac{1}{z_T}\frac{d^2z_T}{d\eta^2}\right)v_T(k,\eta) = 0\, ,
\label{tensor-MS-equation}
\end{align}
with $v_T(k,\eta)$ is the Fourier transformed quantity of the tensor MS variable defined by $\left(v_T\right)_{ij} = z_Th_{ij}$. 
It is evident that both the tensor polarization modes obey the same dynamical equation, for which, we omit the polarization index from the tensor MS variable 
and finally we multiply by a factor of 2 in the tensor power spectrum. 
This is however expected in the present context as there is no parity violating term in the gravitational action. 
This makes the underlying theory different compared to those where the gravitational action contains a parity violating term, from the perspective of gravitational wave. 
Coming back to the present model, considering the Bunch-Davies initial condition for $v_T(k,\eta)$, we solve Eq.~(\ref{tensor-MS-equation}) and given by,
\begin{eqnarray}
v_T(k,\eta) = \frac{\sqrt{\pi\left|\eta\right|}}{2} H_{\theta}^{(2)}(k\left|\eta\right|)
\label{tensor-MS-solution}
\end{eqnarray}
with $\theta = \sqrt{\sigma_T + \frac{1}{4}}$ and $H_{\theta}^{(2)}(k\left|\eta\right|)$ represents the Hermite function of second kind having order $\theta$. 
Consequently the tensor power spectrum for $k$-th mode in the superhorizon scale, i.e., for $k|\eta| \ll 1$, turns out to be,
\begin{align}
\mathcal{P}_{T}(k,\tau) = 2\left[\frac{1}{2\pi}\frac{1}{z_T\left|\eta\right|}\frac{\Gamma(\theta)}{\Gamma(3/2)}\right]^2 \left(\frac{k|\eta|}{2}
\right)^{3 - 2\theta}\, ,
\label{tensor-power-spectrum}
\end{align}
where the multiplication of the factor $2$ is due to the reason that both the tensor polarization modes equally contribute to the 
tensor power spectrum.

\begin{figure}[!h]
\begin{center}
\centering
\includegraphics[width=3.0in,height=3.0in]{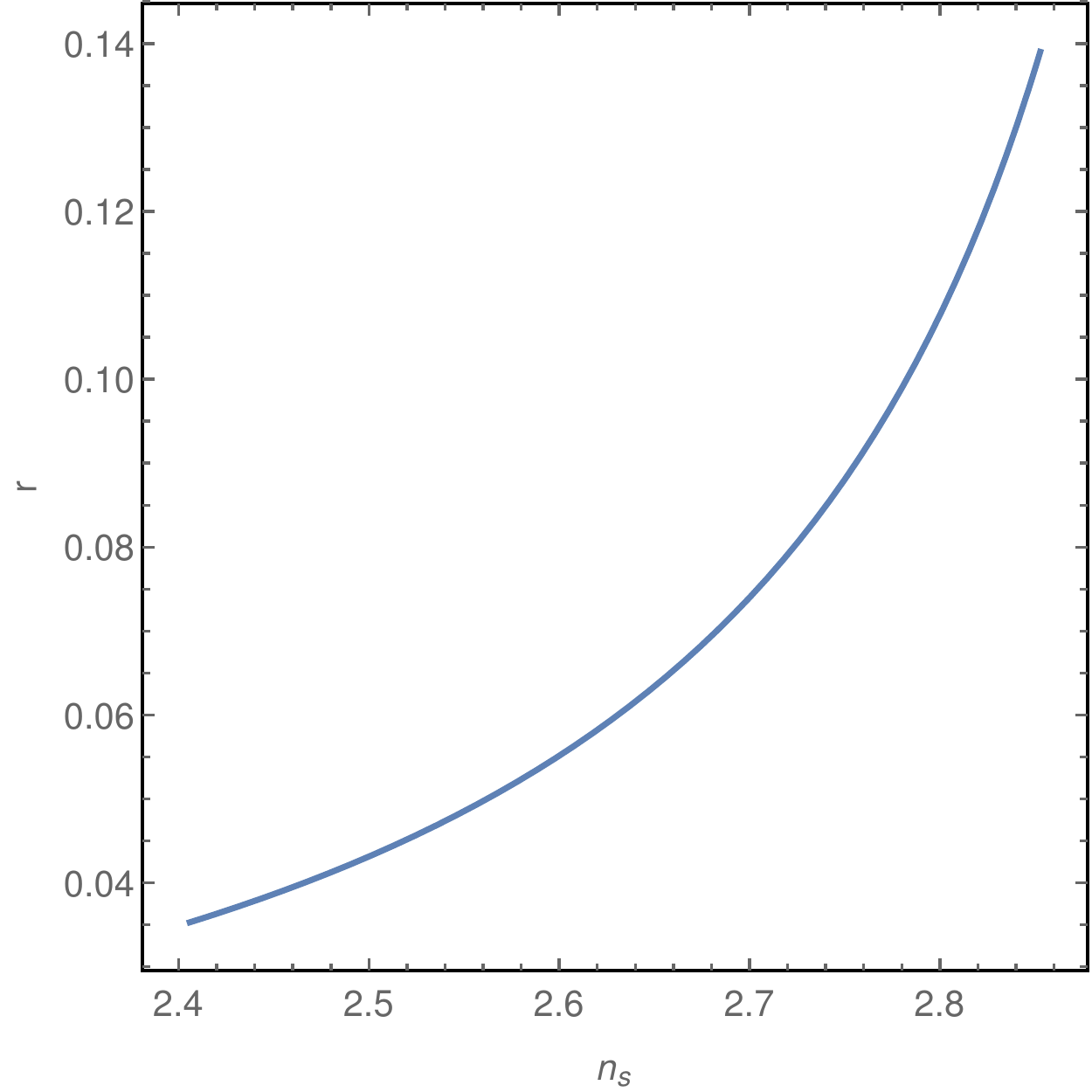}
\caption{Parametric plot of $n_s$ (along $x$-axis) and $r$ (along $y$-axis) 
with respect to $n$ for $0 < n < 1/6$ and $R_0 = 1\,\mathrm{By}^{-2}$. The figure indicates that the tensor-to-scalar ratio satisfies $r > 0.035$ 
which is within the Planck constraints, however the scalar spectral index seems to be greater than the value $2.40$ 
which clearly lies outside the Planck data.}
\label{plot-observable1}
\end{center}
\end{figure}

We now confront the model with the Planck constraints by calculating observable quantities like scalar spectral index, tensor-to-scalar ratio defined by,
\begin{eqnarray}
n_s = 1 + \left. \frac{\partial \ln{\mathcal{P}_{\Psi}}}{\partial \ln{k}} \right|_{h} \, , \quad r=\mathcal{P}_T/\mathcal{P}_{\Psi}\, ,
\label{obs-1}
\end{eqnarray}
and they are constrained by,
\begin{eqnarray}
n_s = 0.9649 \pm 0.0042 \quad \mbox{and} \quad r < 0.064 \, ,
\label{observable-Planck constraint}
\end{eqnarray}
respectively \cite{Akrami:2018odb}. 
The suffix `$h$' in Eq.~(\ref{obs-1}) indicates the horizon crossing instant of the large scale modes around which we will calculate $n_s$ and $r$. 
The expressions of scalar and tensor power spectra in Eq.~(\ref{scalar-power-spectrum-superhorizon}) and 
Eq.~(\ref{tensor-power-spectrum}) immediately lead to the explicit form of $n_s$ and $r$ in the present context as,
\begin{eqnarray}
n_s = 4 - \sqrt{1 + 4\sigma_h} \, , \quad r = 2\left[\frac{z(\eta_h)}{z_T(\eta_h)}\frac{\Gamma(\theta)}{\Gamma(\nu)}\right]^2
\left( k\left|\eta_h\right| \right)^{2(\nu-\theta)}\, ,
\label{obs-2}
\end{eqnarray}
where all the quantities are evaluated at horizon crossing of large scale modes, in particular,
\begin{align}
\nu=&\sqrt{\sigma_h + \frac{1}{4}}\, ; \quad \sigma_h = \xi(\xi - 1)\left[1 - \frac{16n(1-4n^2)}{(1-4n)}
\left(\frac{\widetilde{R}_h}{R_0}\right)^{\frac{1}{2} - n}\right]\, ,\nonumber\\
\theta=&\sqrt{\sigma_{T,h} + \frac{1}{4}} \, ; \quad \sigma_{T,h} = \xi(\xi - 1)\left[1 + \frac{32n(1-4n^2)}{(1-4n)}
\left(\frac{\widetilde{R}_h}{R_0}\right)^{\frac{1}{2} - n}\right]\, ,\nonumber\\
z(\eta_h)=&-\frac{1}{\sqrt{n}}\left(\frac{a_0^n}{\kappa\widetilde{R}_h^{n}}\right)\left[1 - 8n(1+2n)
\left(\frac{\widetilde{R}_h}{R_0}\right)^{\frac{1}{2} - n}\right]\, ,\nonumber\\
z_T(\eta_h)=&\left(\frac{a_0^n}{\kappa\widetilde{R}_h^{n}}\right)\left[1 + 16n(1+2n)
\left(\frac{\widetilde{R}_h}{R_0}\right)^{\frac{1}{2} - n}\right]\, .
\label{obs-3}
\end{align}
Therefore the above expressions contain $\widetilde{R}_h$ that represents the Ricci scalar at the instant of horizon crossing and thus from 
Eq.~(\ref{ricci-scalar-conformal-time}), we can write,
\begin{eqnarray}
\widetilde{R}_h = \left[\frac{1}{a_0^n(1-2n)\left|\eta_h\right|}\right]^{2/(1-2n)}\, ,
\label{obs-4}
\end{eqnarray}
with $\eta_h$ being given from Eq.~(\ref{hc-1}), i.e., 
\begin{eqnarray}
\left|\eta_h\right| = \left(\frac{2n}{1-2n}\right)\frac{1}{k_{CMB}} \approx \left(\frac{2n}{1-2n}\right)\times10\,\mathrm{By}\, .
\label{obs-5}
\end{eqnarray}
In the above estimation, we use $k_{CMB} = 0.002\mathrm{Mpc}^{-1}$ and the conversion $1\,\mathrm{By}^{-1} = \left(\frac{3.17}{1.52}\right)\times10^{-41} \mathrm{GeV}$ may be useful. 
From the above two equations, we get,
\begin{eqnarray}
\widetilde{R}_h = \left[\frac{1}{20na_0^n}\right]^{2/(1-2n)}\, ,
\label{obs-6}
\end{eqnarray}
which presents the $\widetilde{R}_h$ in terms of $a_0$ and $n$. 
As a result, both the scalar spectral index and the tensor-to-scalar ratio of Eq.~(\ref{obs-2}) depend on the parameters $a_0$ and $n$, respectively. 
We take $a_0 = 0.35$ (recall, this value of $a_0$ leads to the central value of the dark energy density parameter $\Omega_D$ in respect to the Planck data, as demonstrated 
in Sec.[\ref{sec-constraints}]). 
In effect, the $n_s$ and $r$ depend only on $n$ which lies within $n < 1/6$ from the condition $C1$. 
Keeping this in mind, we give the parametric plot of $n_s$ and $r$ for $0 < n < 1/6$, see Fig.~[\ref{plot-observable1}]. 
The figure clearly demonstrates that the tensor-to-scalar ratio satisfies $r > 0.035$ and thus remains within the Planck constraint for a suitable range of the parameter $n$. 
However, in regard to the scalar spectral index, it lies $n_s > 2.40$, which in turn indicates a blue tilted scalar power spectrum and thus is not consistent 
with the Planck 2018 constraints. 
Therefore the present model seems not to lead to the simultaneous compatibility of $n_s$ and $r$ in respect to the latest Planck data.

\section{A pre-ekpyrotic phase and viability of the model}
\begin{figure}[!h]
\begin{center}
\centering
\includegraphics[width=3.0in,height=3.0in]{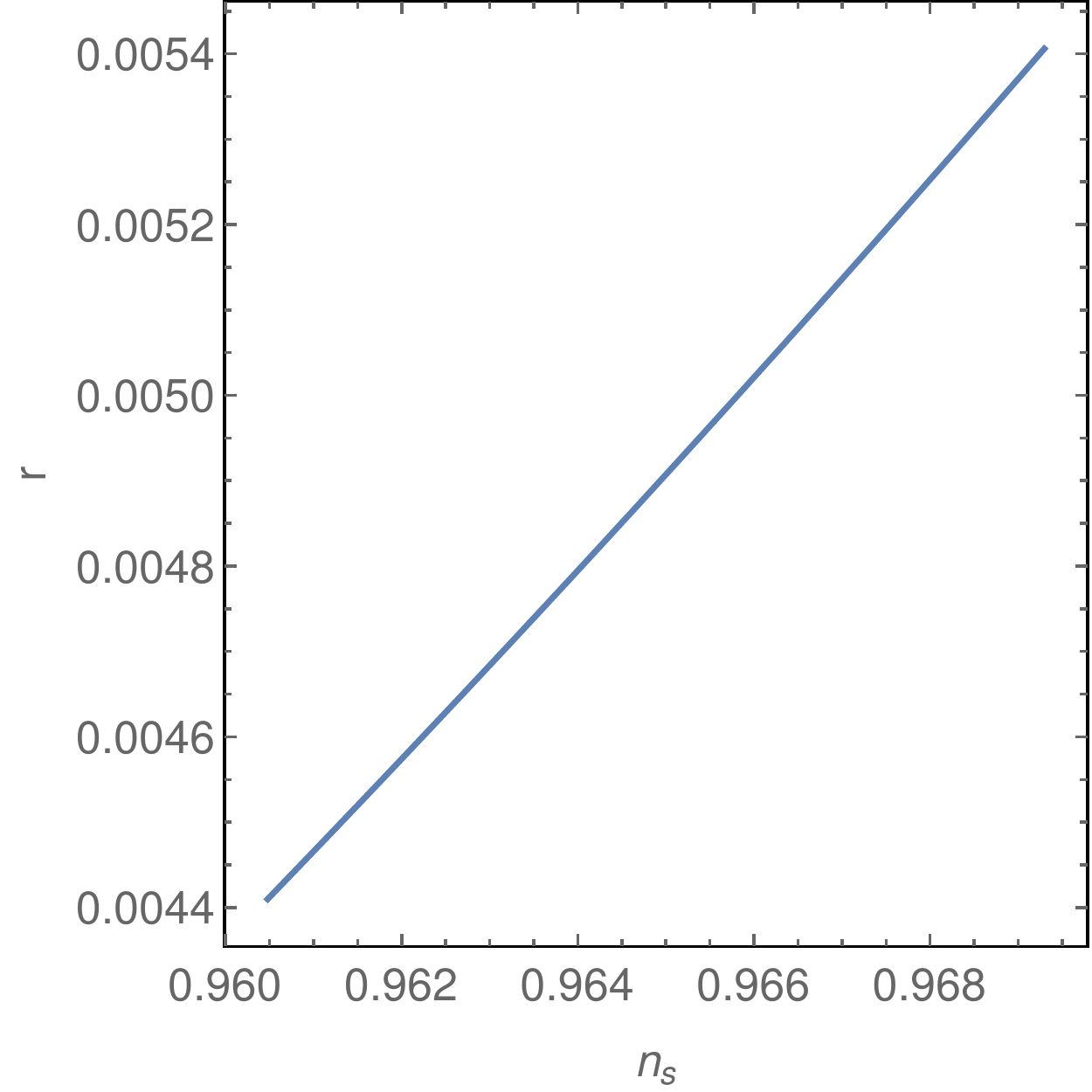}
\caption{Parametric plot of $n_s$ (along $x$-axis) and $r$ (along $y$-axis) 
with respect to $m$ for $0.2544 \lesssim m \lesssim 0.2547$ and $R_0 = 1\,\mathrm{By}^{-2}$. 
The figure clearly demonstrates that the scalar spectral index and tensor to 
scalar ratio, on scales that cross the horizon during the pre-ekpyrotic stage, are 
simultaneously comptible with the Planck data within the parametric regime $0.2544 \lesssim m \lesssim 0.2547$.}
\label{plot-observable2}
\end{center}
\end{figure}

We just verified that the cosmological scenario, where the universe is dominated by an ekpyrotic phase of contraction before the bounce 
in the context of ghost free $f(R,\mathcal{G})$ gravity theory, leads to a blue tilted scalar spectrum which is not consistent with the Planck results. 
Such an inconsistency arises due to the fact that the large scale modes cross the horizon during the ekpyrotic phase where $a(t) \sim t^{2n}$ with $n < 1/6$. 
This result is in agreement with \cite{Cai:2012va,Cai:2014zga} where the authors studied an ekpyrotic bounce scenario in a single scalar field 
model with non-trivial scalar potential and non-standard kinetic term without/with accounting loop quantum effects around the bounce, unlike to the present 
$f(R,\mathcal{G})$ gravity theory. 
In attempt to get a scale invariant scalar power spectrum, we consider a pre-ekpyrotic phase where the scale factor behaves 
as $a_p(t) \sim t^{2m}$ with $m > 1/6$, in particular, the pre-ekpyrotic phase is described by (\cite{Cai:2012va,Cai:2014zga}),
\begin{eqnarray}
a_p(t) = a_e^{m}\left(\frac{t- T}{t_e - T}\right)^{2m} \quad \mbox{with} \quad m>1/6\, ,
\label{pre-scale factor}
\end{eqnarray}
and re-examine the scalar power spectrum. 
In the above expression, $t_e$ is the instant when the transition from pre-ekpyrotic to ekpyrotic contraction occurs and $T$ is a fiducial time to make everything consistent. 
Moreover for $m = 1/3$, $a_p(t)$ describes a matter dominated epoch. 
Thus in this modified cosmological scenario, the scale factor of the universe is described by:
\begin{align}
a_p(t) = a_e^{m}\left(\frac{t- T}{t_e - T}\right)^{2m} \quad &\mbox{with} \quad m>1/6\, ,  \quad \mathrm{for}~t \leq t_e\, ,\nonumber\\
a(t) = \left[1 + a_0t^2\right]^n\times \exp{\left[\frac{1}{(\alpha-1)}\left( t_s - t \right)^{1-\alpha}\right]} \quad &\mbox{with} \quad n< 1/6 \, , \quad \mathrm{for}~t \geq t_e\, .
\label{full-scale-factor}
\end{align} 
The continuity relations of scale factor and Hubble parameter at the transition time $t = t_e$ are,
\begin{align}
a_e^m=&\left(1 + a_0t_e^2\right)^n \exp{\left[\frac{\left(t_s - t_e\right)^{1-\alpha}}{(\alpha - 1)}\right]}\, ,\nonumber\\
\frac{2m}{t_e - T}=&\frac{2a_0nt_e}{\left(1 + a_0t_e^2\right)} + \frac{1}{\left(t_s - t_e\right)^{\alpha}}\, ,
\label{continuity}
\end{align}
respectively. 
In presence of such a pre-ekpyrotic phase, the large scale modes cross the horizon either during the pre-ekpyrotic stage or during the 
ekpyrotic stage depending on whether the transition time ($t_e$) is larger or less than the horizon crossing instant of large scale modes. 
We need the first case, i.e., the large scale modes cross the horizon during the pre-ekpyrotic stage, with a hope for a scale invariant power spectrum. 
In order to do so, the transition time should be larger than the horizon crossing instant, i.e., $t_e > t_h$ or equivalently $\left|t_e\right| < \left|t_h\right|$. 
The conformal time for the scale factor $a_p(t)$ comes as,
\begin{eqnarray}
\eta_p(t) = \int\frac{dt}{a_p(t)} = \left[\frac{\left(t_e - T\right)^{2m}}{a_e^{m}(1 - 2m)}\right]\left(t - T\right)^{1-2m}\, ,
\label{conformal-pre-ekpyrotic}
\end{eqnarray}
and consequently, the horizon crossing time for $k$-th mode is given by,
\begin{eqnarray}
\left|\eta_h\right| = \left(\frac{2m}{1-2m}\right)\frac{1}{k}\, .
\label{hc-1-pre-ekpyrotic}
\end{eqnarray}
The above expression of $\eta_h$ has same form as of Eq.~(\ref{hc-1}) with the replacement of the parameter $n$ to $m$ -- this 
is, however, expected from the comparison between the scale factors during pre-ekpyrotic and ekpyrotic phases, respectively. 
Therefore the horizon crossing instant for $k = 0.002\mathrm{Mpc}^{-1}$ is obtained as,
\begin{eqnarray}
\left|\eta_h\right| \approx \left(\frac{2m}{1-2m}\right)\times10\,\mathrm{By}\, .
\label{hc-2-pre-ekpyrotic}
\end{eqnarray}
Thus we may argue that the transition time $\left|t_e\right| < 10\,\mathrm{By}$ in order to cross the horizon of the large scale modes during the pre-ekpyrotic phase. 
Following the same procedure as of Sec.[\ref{sec-perturbation}], we determine the scalar spectral index and tensor to scalar ratio in the modified 
scenario where the ekpyrotic phase is preceded by a period of a pre-ekpyrotic contraction:
\begin{eqnarray}
n_s = 4 - \sqrt{1 + 4\sigma_h} \, , \quad r = 2\left[\frac{z(\eta_h)}{z_T(\eta_h)}\frac{\Gamma(\theta)}{\Gamma(\nu)}\right]^2
\left(\frac{2m}{1-2m}\right)^{2(\nu-\theta)}\, ,
\label{obs-1-pre}
\end{eqnarray}
with the quantities given by,
\begin{align}
\nu =&\sqrt{\sigma_h + \frac{1}{4}}\, ; \quad \sigma_h = \frac{2m\left(4m-1\right)}{(1-2m)^2}\left[1 - \frac{16m(1-4m^2)}{(1-4m)}
\left(\frac{\widetilde{R}_h}{R_0}\right)^{\frac{1}{2} - m}\right]\, ,\nonumber\\
\theta =&\sqrt{\sigma_{T,h} + \frac{1}{4}} \, ; \quad \sigma_{T,h} = \frac{2m\left(4m-1\right)}{(1-2m)^2}\left[1 + \frac{32m(1-4m^2)}{(1-4m)}
\left(\frac{\widetilde{R}_h}{R_0}\right)^{\frac{1}{2} - m}\right]\, ,\nonumber\\
\frac{z(\eta_h)}{z_T(\eta_h)} =&\frac{1}{\sqrt{m}}\left[\frac{1-8m(1+2m)\left(\widetilde{R}_h/R_0\right)^{\frac{1}{2} - m}}
{1 + 16m(1+2m)\left(\widetilde{R}_h/R_0\right)^{\frac{1}{2} - m}}\right]\, .
\label{obs-2-pre}
\end{align}
As mentioned earlier, due to $\left|t_e\right| < 10\,\mathrm{By}$, the large scale modes cross the horizon during the pre-ekpyrotic stage and thus 
the quantity $\widetilde{R}_h$ (i.e., the Ricci scalar at the horizon crossing instant) can be determined by using the pre-ekpyrotic scale factor 
$a_p(t)$ in Eq.~(\ref{pre-scale factor}). In effect, $\widetilde{R}_h$ is given by,
\begin{eqnarray}
\widetilde{R}_h = \left[\frac{\left(t_e - T\right)^{2m}}{20ma_e^m}\right]^{2/(1-2m)}\, .
\label{ricci-pre-1}
\end{eqnarray}
Therefore to determine $\widetilde{R}_h$, we need $a_e$ and $(t_e - T)$, and for this purpose, we use the continuity relations given in 
Eq.~(\ref{continuity}). With $a_0 = 0.35$, $\omega = 3$, $t_s = 30\,\mathrm{By}$ and $\alpha = 4/3$ (the same 
parameter values used throughout the paper), Eq.~(\ref{continuity}) immediately leads to $a_e$ and $(t_e - T)$ in terms of the exponent $m$ as, 
\begin{eqnarray}
a_e = \left(3.163\right)^m \quad \mbox{and} \quad \left(t_e - T\right) = 96m\, \mathrm{By}\, ,
\end{eqnarray}
respectively. 
The above expressions along with Eq.~(\ref{obs-1-pre}) clearly indicate that both the scalar spectral index and tensor to scalar ratio in such modified scenario depend only on $m$. 
As a result, we give the simultaneous plot of $n_s$ and $r$ with respect to $m$, see Fig.~[\ref{plot-observable2}]. 
The figure clearly demonstrates that the scalar spectral index and tensor to scalar ration, on scales that cross the horizon during the pre-ekpyrotic stage 
of contraction, are simultaneously compatible with the Planck data within the parametric regime $0.2544 \lesssim m \lesssim 0.2547$. 
Therefore in the present $f(R,\mathcal{G})$ cosmological scenario 
where an ekpyrotic bounce is smoothly unified to the dark energy era, the existence of a pre-ekpyrotic phase 
with $0.2544 \lesssim m \lesssim 0.2547$ makes the observable quantities at large scale modes compatible with the latest Planck data. 
Furthermore, the cosmic horizon crossing time of the large scale mode (in particular, $k = 0.002\mathrm{Mpc}^{-1}$) during the pre-ekpyrotic phase 
comes as $\left|t_h\right| = \left(\frac{20ma_e^m}{\left(t_e - T\right)^{2m}}\right)^{1/(1-2m)}$ which, 
due to $0.2544 \lesssim m \lesssim 0.2547$, lies within $10.446\,\mathrm{By} \lesssim \left|t_h\right| \lesssim 10.449\,\mathrm{By}$. 

At this point, the importance of the Gauss-Bonnet (GB) coupling function deserves to mention. Recall, that the GB coupling function 
is connected to the constant $R_0$ via $\kappa^2h_0 = \left(1/R_0\right)^{\frac{1}{2} - n}$, see Eq.~(\ref{h0-dimension}). 
Therefore, in absence of the GB coupling function, i.e., 
for $h_0 = 0$ or equivalently $1/R_0 = 0$, the scalar spectral index and the tensor to scalar ratio from Eq.~(\ref{obs-1-pre}) become 
$n_s = \left(5-14m\right)/\left(1-2m\right)$ and $r = 2/m$, respectively. 
Such expressions of $n_s$ and $r$ clearly indicates that the scalar power spectrum gets scale invariant (i.e., $n_s = 1$) for $m= 1/3$ leading to a matter dominated epoch 
before the ekpyrotic phase, however the corresponding tensor to scalar ratio becomes $r = 6$ which is far larger than the Planck data. 
Therefore in the absence of GB coupling function, although the scalar power spectrum is found to be 
scale invariant for an early matter dominated epoch, but at the same time, it predicts a large tensor to scalar ratio. 
This is however expected, because for $h_0 = 0$, the action of the model resembles with a scalar-tensor action which indeed leads 
to a scale invariant scalar power spectrum with a large tensor to scalar ratio in matter bounce scenario. 
On contrary, in the present context of $f(R,\mathcal{G})$ gravity compatible with GW170817, the presence of the GB coupling function considerably affects the observable 
quantities (see Eq.~(\ref{obs-1-pre}), and as a result, both the $n_s$ and $r$ seem to be consistent with the Planck data for a suitable pre-ekpyrotic 
stage satisfying $0.2544 \lesssim m \lesssim 0.2547$. 
Actually the GB coupling function enhances the amplitude of tensor perturbation and consequently reduces the tensor 
to scalar ratio compared to the case where the GB coupling function is absent. 
This clearly demonstrates the importance of $f(R,\mathcal{G})$ gravity in making the present bounce phenomenology consistent with the observational data. 

\section{Conclusion}
In this paper, we propose a non-singular cosmological scenario in a ghost free $f(R,\mathcal{G})$ model, where the universe initially contracts through 
an ekpyrotic phase with a bouncing like behaviour, and accordingly after it bounces off, it smoothly transits to a radiation or matter like 
deceleration era which is further smoothly connected to the dark energy era at current epoch. 
The ghost free nature of the model is ensured by the presence of the Lagrange multiplier in the action, as developed in \cite{Nojiri:2018ouv}. 
We consider the Gauss-Bonnet coupling function ($h(t)$) in such a way that the speed of gravitational wave becomes unity and the model gets compatible with the striking event GW170817. 
This happens for a class of GB coupling function which satisfies a constraint equation of the form $\ddot{h} = \dot{h}H$ (where $H$ is the Hubble parameter of the universe). 
By imposing this condition and using several reconstruction technique, we obtain the explicit form of scalar field potential as well as the GB coupling function, 
which seems to source a smooth unified from an ekpyrotic bounce to the dark energy era with an intermediate deceleration stage. 
The existence of ekpyrotic phase resolves the anisotropic problem from the background evolution and leads to a natural bounce scenario that is free from the BKL instability. 
Consequently the effective equation of state parameter is determined, which ensures the successive transitions of the universe from acceleration-to-deceleration era or vice-versa. 
In regard to the dark energy era, the theoretical predictions of various observable quantities (like dark energy EoS parameter, 
dark energy density parameter etc.) are confronted with Planck+SNe+BAO data, and it turns out that the dark energy parameters are indeed 
compatible with the observational data for suitable parametric regime of the model. 
The Hubble radius shows an asymmetric evolution around the bounce, in particular, it increases with time and diverges to infinity at late contracting stage, 
however, due to the presence of dark energy era, it decreases near about the current epoch of universe (i.e., around $t = 13.8\,\mathrm{By}$). 
Such evolution of the Hubble radius leads to the generation era of primordial perturbation modes far before the bounce at the deep sub-Hubble regime. 
Accordingly we perform the scalar and tensor perturbation in the present cosmological scenario. 
As a result, the scalar power spectrum is found to be highly blue tilted at large scale modes, which is not consistent with the Planck constraints. 
Such inconsistency occurs due to the fact that the large scale modes cross the horizon during the ekpyrotic phase of contraction. 
Therefore a modified scenario is proposed, where a stage of pre-ekpyrotic contraction is considered, with the equation of state parameter being less than unity. 
The transition from the pre-ekpyrotic to the ekpyrotic stage is shown to be a smooth transition by demanding the continuity of scale factor and Hubble parameter at the transition time. 
It turns out that in the extended scenario where the ekpyrotic phase is preceded by a stage of pre-ekpyrotic contraction, both the scalar spectral index 
and tensor to scalar ratio at large scales (scales which exit the Hubble horizon during the pre-ekpyrotic stage) behave according to the Planck data. 
In this regard, the GB coupling function is found to have considerable effects in making the primordial quantities consistent with the observational data. 
In absence of the GB coupling function, although the scalar power spectrum becomes scale invariant for an early matter dominated epoch before the ekpyrotic phase, 
but it predicts a large tensor to scalar ratio -- similar to a scalar-tensor bounce model. 
However, in the present context of $f(R,\mathcal{G})$ model, the GB coupling significantly modifies the perturbation power spectra, and as a result, 
the scalar spectral index and tensor to scalar ratio at large scales (that exit the Hubble radius during the pre-ekpyrotic stage) 
get simultaneously compatible with the latest Planck data for a suitable pre-ekpyrotic phase described by the scale factor: $a(t) \sim t^{2m}$ with 
the exponent satisfying $0.2544 \lesssim m \lesssim 0.2547$. Actually the GB coupling suppresses the amplitude of tensor perturbation and 
consequently reduces the tensor to scalar ratio compared to the case where the GB coupling is absent. 
This clearly argues the importance of $f(R,\mathcal{G})$ gravity in making a viable phenomenology of the present cosmological scenario.

In summary, the present model provides a smooth unified cosmological scenario from a non-singular ekpyrotic bounce to the dark
energy epoch with an intermediate deceleration era in the ghost free $f(R,\mathcal{G})$ theory compatible with GW170817, 
where the Gauss-Bonnet coupling plays a crucial role in reducing the tensor to scalar ratio and makes it consistent with the Planck results. 
Therefore the proposed bounce scenario is able to concomitantly address the issues like -- (1) explaining the current dark energy epoch, (2) avoiding the 
BKL instability and (3) predicting a viable scalar spectral tilt and tensor to scalar ratio, respectively.

\section*{Acknowledgments}
This work was supported in part  by the JSPS Grant-in-Aid for Scientific Research (C)
No. 18K03615 (SN) and by MINECO (Spain), project PID2019-104397GB-I00 (SDO). This work was partially supported by the program Unidad de Excelencia
MarÃ­a de Maeztu CEX2020-001058-M. This research was also supported in part by the 
International Centre for Theoretical Sciences (ICTS) for the online program - Physics of the Early Universe (code: ICTS/peu2022/1). 
TP thanks L. Sriramkumar for useful discussions.

\end{document}